\documentclass[aps,prl,final,twocolumn,superscriptaddress,floatfix,showpacs,10pt]{revtex4-1}

\usepackage[latin9]{inputenc}
\usepackage{graphicx,amssymb,soul,color,amsmath,bm}
\usepackage{epstopdf,hyperref}
\usepackage{times}

\hypersetup{colorlinks=true,citecolor=blue}

\begin{document}

\title{
Exotic Magnetic Order in the Orbital-Selective Mott Regime of Multiorbital Systems
}

\author{Juli\'an Rinc\'on}
\affiliation{Center for Nanophase Materials Sciences, Oak Ridge National Laboratory, Oak Ridge, Tennessee 37831, USA}
\affiliation{Materials Science and Technology Division, Oak Ridge National Laboratory, Oak Ridge, Tennessee 37831, USA}

\author{Adriana Moreo}
\affiliation{Materials Science and Technology Division, Oak Ridge National Laboratory, Oak Ridge, Tennessee 37831, USA}
\affiliation{Department of Physics and Astronomy, The University of Tennessee, Knoxville, Tennessee 37996, USA}

\author{Gonzalo Alvarez}
\affiliation{Center for Nanophase Materials Sciences, Oak Ridge National Laboratory, Oak Ridge, Tennessee 37831, USA}
\affiliation{Computer Science \& Mathematics Division, Oak Ridge National Laboratory, Oak Ridge, Tennessee 37831, USA}

\author{Elbio Dagotto}
\affiliation{Materials Science and Technology Division, Oak Ridge National Laboratory, Oak Ridge, Tennessee 37831, USA}
\affiliation{Department of Physics and Astronomy, The University of Tennessee, Knoxville, Tennessee 37996, USA}

\date{\today}

\begin{abstract}
The orbital-selective Mott phase (OSMP) of  multiorbital Hubbard models 
has been extensively analyzed before using static
and dynamical mean-field approximations. In parallel, 
the properties of Block states (antiferromagnetically coupled ferromagnetic spin clusters) 
in Fe-based superconductors have also been much
discussed. The present effort uses numerically exact techniques in one-dimensional systems to 
report the observation of Block states within the OSMP regime, 
connecting two seemingly independent areas of research, and providing 
analogies with the physics of Double-Exchange models.
\end{abstract}

\pacs{71.30.+h, 71.27.+a, 71.10.Fd, 71.10.Fd}

\maketitle

{\it Introduction.} The combined interplay of charge, spin, lattice, and orbital degrees 
of freedom have led to an enormous variety of emergent phenomena in strongly correlated 
systems. A prototypical example is the half-filled single-orbital metal-insulator transition, 
that is realized in materials such as La$_2$CuO$_4$, a parent compound
of the Cu-based high temperature superconductors.
If several active orbitals are also considered in the study of this transition, an even richer 
phase diagram is anticipated, where states such as band insulators, correlated 
metals, and orbital-selective Mott phases (OSMP) can be stabilized. In particular, 
the study of the OSMP and its associated orbital-selective Mott transition 
has attracted considerable attention in recent years~\cite{Anisimov,Vojta,Georges13}.

The OSMP is a state 
where even though Mott insulator (MI) physics 
occurs, it is restricted to 
a subset of all the active orbitals present in the problem.
This state has narrow-band localized electrons related 
to the MI orbitals, coexisting with wide-band itinerant electrons 
at the other orbitals~\cite{Liebsch04,Biermann05,Medici09}.  
To stabilize the OSMP, a robust Hund interaction $J$ is needed.
In general, the hybridization within 
orbitals $\gamma$, $V_{\gamma,\gamma'}$, and crystal fields, $\Delta_{\gamma}$, 
work against $J$ since they favor low-spin ground states. 
Therefore, if $J\gg \Delta_{\gamma},\, V_{\gamma,\gamma'}$, 
the OSMP is expected to be stable and display robust local moments~\cite{Medici09}.

Several studies focused on the effects 
of interactions, filling fractions, etc., on the stability
of orbital-selective phases~\cite{Georges13}. 
This previous theoretical work was performed within mean-field approximations 
(such as Dynamical Mean Field Theory~\cite{Liebsch04,Biermann05,Medici09,Ishida10,Medici11b}, 
slave-spins~\cite{Medici09,Medici11b,Yu12,Yu13,Medici-new}, 
or Hartree-Fock~\cite{Bascones12,Valenzuela13}). 
Using these methods the OSMP stability conditions have been established.
However, to our knowledge, there have been no detailed studies 
of the influence of full quantum fluctuations on this phase
and therefore, and more importantly for our purposes, 
of their low-temperature electronic and magnetic properties. 

Recently, these issues 
received considerable attention in the Fe-based superconductors community.
In this context, multiorbital models containing 
Hubbard $U$ and $J$ interactions, 
as well as crystal-field splittings,
are widely employed, and the 
existence of OSMP regimes has been extensively 
investigated~\cite{Medici09,Ishida10,Yu12,Yu13,Bascones12,Valenzuela13,Liebsch11,Yi-2012,lanata,Greger13}.
For these reasons, our models 
are chosen to resemble qualitatively models for pnictides and selenides, 
so that our conclusions could be of potential relevance in that context.

Our aim is two-fold:~$(i)$ By employing
techniques beyond mean-field to study the phase diagrams of 
three-orbital Hubbard models, the robustness of the OSMP
to full quantum fluctuations 
can be confirmed via numerical simulations. 
$(ii)$ More importantly, once the stability of this state has been 
established, its charge and magnetic orders can be explored. 
Our main result is that full ferromagnetic (FM)
and exotic ``Block'' states have been found
to be stable within the OSMP regime. Block phases, such as the  antiferromagnetic (AFM)
state made of 2$\times$2 FM-clusters, were discussed extensively 
in the selenides
literature mainly in the presence of iron 
vacancies~\cite{bao,cao,luo-vacancies,Yu11,huang-mou,dong,BNL,RMP,Dai} 
or in special geometries 
such as two-leg ladders~\cite{caron1,caron2,luo-ladders}, but not in the OSMP framework. 
Our new results suggest that the concepts of OSMP and Block phases, until now
separately considered, are actually related after showing 
that Block states are stabilized {\it within} the OSMP regime.

Our study is performed in a one-dimensional geometry. 
This restriction arises from the need 
to employ accurate techniques such
as the Density Matrix Renormalization Group 
(DMRG)~\cite{dmrg1,dmrg2,dmrg3}. Thus, our conclusions are
only {\it suggestive} of similar physics in 
the layered Fe-based superconductors, and
only further work can confirm this assumption. 
However, there are real quasi one-dimensional
materials, such as the previously mentioned 
ladders~\cite{caron1,caron2,luo-ladders}, that
may provide a direct physical realization of our results.

{\it Model.}
The Hamiltonians used are multiorbital Hubbard models 
composed of kinetic energy and interacting terms: $H=H_K + H_{\rm int}$. 
The kinetic contribution is written as
\begin{equation}
H_K = -\sum_{i\sigma\gamma\gamma'} t_{\gamma\gamma'} \bigl( c^+_{i \sigma \gamma}c_{i+1 \sigma \gamma'} + \mathrm{H.c.} \bigr) 
+ \sum_{i\sigma\gamma}\Delta_\gamma n_{i \sigma \gamma},
\end{equation}
where $t_{\gamma \gamma'}$ denotes a (symmetric) hopping amplitude 
defined in orbital space $\{\gamma\}$ connecting the lattice 
sites $i$ and $i+1$ $(\gamma=0,\,1,\,2)$ of a one dimensional lattice of length $L$.
The hopping amplitudes used here are 
(eV units): $t_{00} = t_{11} = -0.5$, $t_{22} = -0.15$, $t_{02} = t_{12} = 0.1$, 
and $t_{01} = 0$, with an associated 
total bandwidth $W=4.9|t_{00}|$ (the individual orbital bandwidths 
$W_\gamma/|t_{00}|$ are 3.69, 3.96, and 1.54, for $\gamma = 0, 1,$ and $2$, 
respectively).  
Both hoppings and $W$ are
comparable in magnitude to those used in more realistic
pnictides models~\cite{daghofer}. 
Only hybridizations between orbitals 0 and 2, and 1 and 2, are considered. 
$\Delta_\gamma$ defines a crystal-field splitting which is orbital-dependent with values 
$\Delta_0 = -0.1$, $\Delta_1 = 0$, and $\Delta_2 = 0.8$. 
All these parameters are {\it phenomenological}, i.e.~not derived 
from {\it ab-initio} calculations. Their values were chosen so that the
band structure [shown in the inset of Fig.~2(a)] qualitatively resembles that of higher 
dimensional pnictides, with a hole pocket at $k=0$ and electron pockets at $k=\pm \pi$. 
This model will be referred to as ``Model~1'' while a ``Model~2'' with slightly 
different parameters will be discussed in the Suppl.~Material (SM)~\cite{suppl}.
The Coulombic repulsion interacting portion of the Hamiltonian is
\begin{multline}
H_{\rm int} = U\sum_{i\gamma} n_{i\uparrow\gamma}n_{i\downarrow\gamma} 
+\left(U'-J/2\right)\sum_{i\gamma<\gamma'} n_{i\gamma}n_{i\gamma'} \\
  -2J\sum_{i\gamma<\gamma'} \mathbf{S}_{i\gamma} \cdot \mathbf{S}_{i\gamma'} 
+J\sum_{i\gamma<\gamma'} \left( P^+_{i\gamma} P_{i\gamma'} + \mathrm{H.c.} \right),
\end{multline}
containing the standard intra-orbital Hubbard repulsion, $U,$ and Hund's rule coupling, $J$. For
SU(2) symmetric systems, the relation $U' = U - 2J$ holds. 
$c_{i \sigma \gamma}$ annihilates 
an electron with spin $\sigma$ at orbital $\gamma$ and site $i$, 
and $n_{i \sigma \gamma}$ counts 
electrons at $i$ with quantum numbers $(\sigma,\gamma)$. 
The operator $\mathbf{S}_{i \gamma}$ ($n_{i \gamma}$) is the spin (total 
electronic density) at orbital $\gamma$ and site $i$; and the definition $P_{i\gamma}=c_{i\downarrow\gamma}c_{i\uparrow\gamma}$ was
introduced. The electronic density per orbital 
is fixed to $n=4/3$, i.e., four electrons every three orbitals, 
in analogy with the filling used in the modeling of iron 
superconductors with three orbitals~\cite{daghofer}. 
As many-body technique, 
the DMRG method~\cite{dmrg1,dmrg2,dmrg3} was used, with technical details provided
in the SM~\cite{suppl}.

\begin{figure}
\includegraphics*[width=.85\columnwidth]{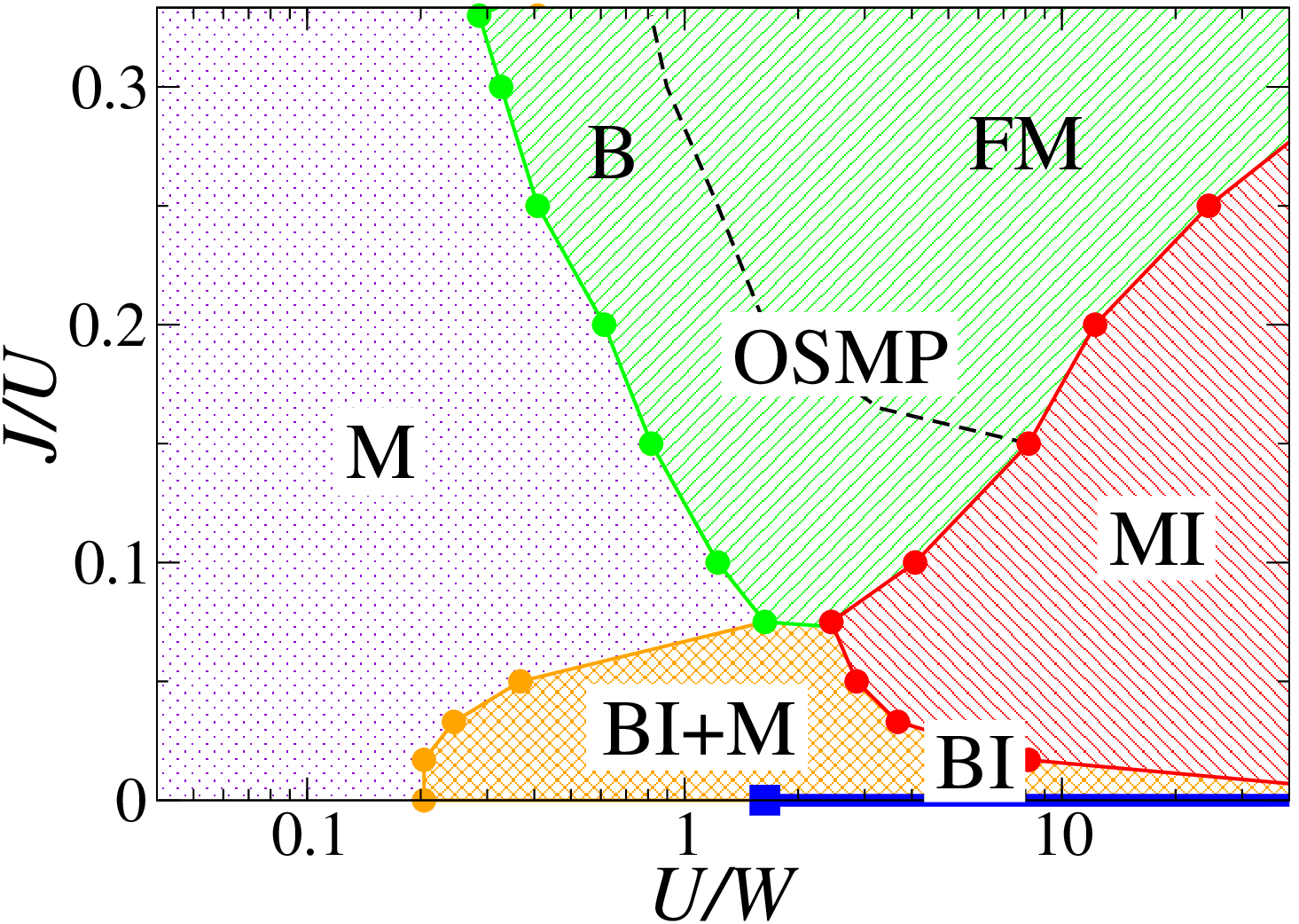}
\caption{Phase diagram of the three-orbital Model~1 for $n = 4/3$. 
The different phases are labeled as: metal (M), band insulator (BI), 
a metallic state resembling the BI state (BI+M), Mott insulator (MI), 
and orbital-selective Mott phase (OSMP). Within the OSMP regime, 
it is possible to distinguish between Block (B) and FM states.}
\label{fig:1}
\end{figure}

\begin{figure}
\includegraphics*[width=.85\columnwidth]{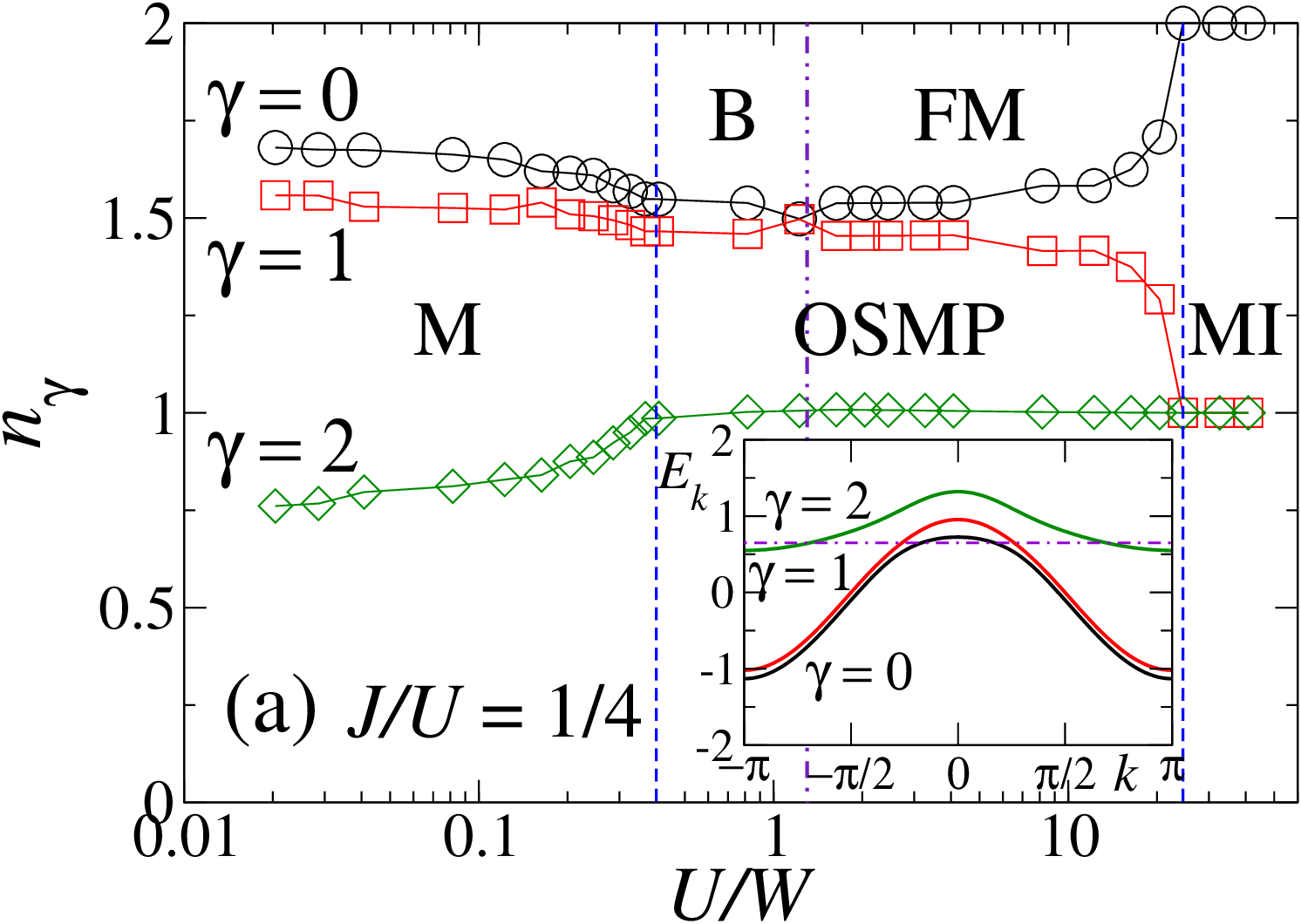}
\includegraphics*[width=.85\columnwidth]{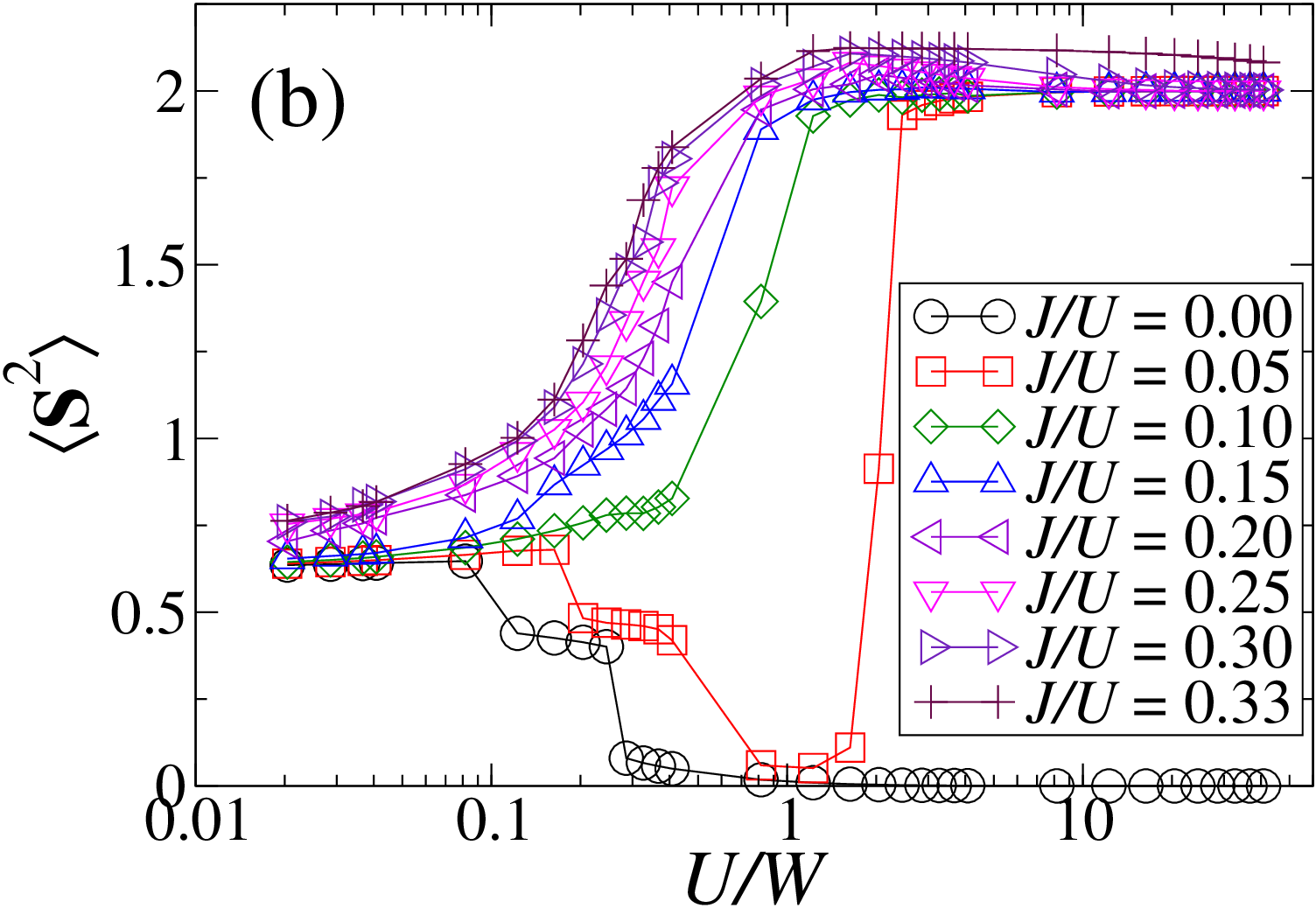}
\caption{Results for Model~1: (a) Electronic occupancy, $n_\gamma$, corresponding to orbital $\gamma$
vs.~$U$ in bandwidth $W$ units, at $J/U = 1/4$, and for $L=24$.
{\it Inset:} Band structure of Model~1 (eV units).  
Violet (dash dotted) line is the $n=4/3$ Fermi level.
(b) Mean value of the square of the total spin at each site, at
several values of $J/U$. Within the OSMP, a robust magnetic moment is observed.}
\label{fig:2}
\end{figure}

{\it Results.} In Fig.~\ref{fig:1}, the phase diagram of Model~1 is shown, 
based on the DMRG measurements of the orbital occupancies $n_\gamma$ and 
the square of the spin operator at every site (see Fig.~\ref{fig:2}). 
Two phases are obvious: a metallic weakly interacting state 
M at small $U$ and a MI regime at large $U,$ where 
$n_0=2$, $n_1=n_2=1$ minimize the double-occupancy 
energy penalization and $J$ induces a spin 1 state at each site (orbital 0 is 
doubly occupied because of the small but nonzero split between orbitals 1 and 0).
Naturally, the spin order is staggered in the MI phase.
Less obvious are the other two phases in Fig.~\ref{fig:1}.
For example, a correlated ``band insulator'' (BI) with 
 $n_0 = n_1 = 2$ and $n_2 = 0$ is found in a region bounded by
$J$ less than the crystal-field splittings $\Delta_{\gamma}$, 
so that the low-spin state is favored, and $U/W$ not too large, so that double
occupancy is not heavily suppressed by $U$. At $J$ exactly 0.0, the BI 
state survives for any value of $U/W$ because in this line $U=U'$. But
at large $U/W$, a tiny $J$ is sufficient to destabilize the BI state into
the MI state.  The related BI+M state 
has $n_0 \sim n_1 \sim 2$ and $n_2 \sim 0$: a metallic state
with characteristics close to the BI~\cite{rong-comment}.
Clearly, increasing $J/U$ the BI/BI+M phases are
suppressed. In fact, with increasing $U/W$ the M state is the most stable 
at $J/U\sim 0.075$ because of the competition $J$ vs. $\Delta_{\gamma}$.
Since the BI/BI+M states are not our main focus,
additional properties are in the SM~\cite{suppl}.

Our most important result in Fig.~\ref{fig:1} is the presence 
of a prominent OSMP regime, stabilized after $J$ becomes larger than a 
threshold that depends on $\Delta_{\gamma}$. 
The OSMP contains the region $J/U \sim 1/4$ at intermediate $U/W$
believed to be realistic~\cite{Haule,luo-neutrons,Kotliar,valenti}.
For the prototypical value $J/U=1/4$,
in the small-$U$ metallic regime the $n_\gamma$ values evolve 
smoothly from the non-interacting limit.
However, at a critical $U/W$, the $\gamma=2$ orbital population
reaches 1 and stays there in a wide window
of couplings, while the other two densities develop a 
value $\sim 1.5$~[Fig.~\ref{fig:2}(a)]. 
These results are robust against changes in $L$ 
and they are compatible with the presence of an OSMP,
that eventually ends at a second critical $U/W$ when 
the transition to the MI regime occurs.
In the SM~\cite{suppl}, results similar to
Fig.~\ref{fig:2} (a) but adding two holes are shown: while 
$n_2$ remains at 1 in the OSMP regime, now
$n_0 \sim n_1 \sim 1.37$, showing that $n_0$ and $n_1$ evolve with doping while $n_2$ is locked as
expected in the OSMP~\cite{footnotedelta}.
The formation of a robust local moment
in the OSMP is shown in Fig.~\ref{fig:2}(b). These DMRG 
results confirm the presence of the OSMP
even with full quantum fluctuations incorporated. The OSMP
regime is eventually suppressed by decreasing $J/U$: at
weak $U/W$ coupling because of the competition with $\Delta_\gamma$ 
that favors the M and BI+M states, 
and at strong $U/W$ coupling because of the competition with the MI state. 

\begin{figure}
\includegraphics*[width=\columnwidth]{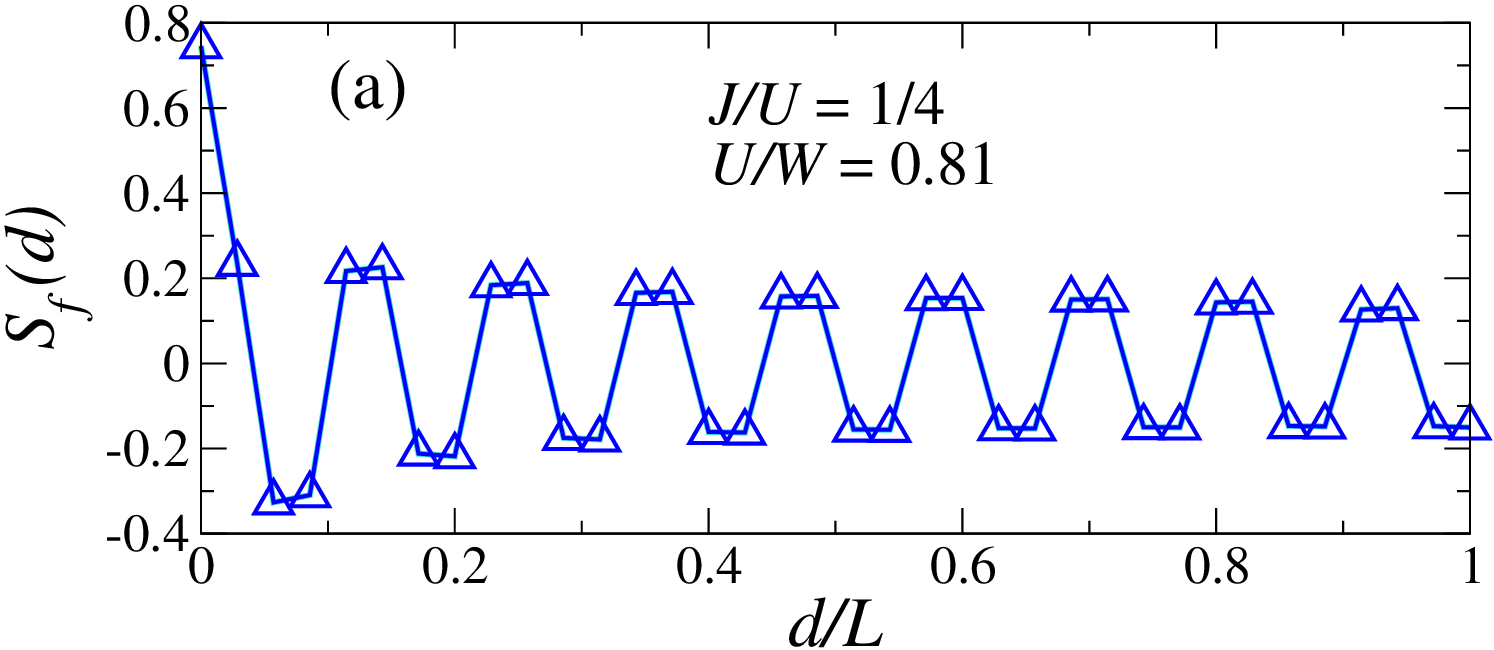}
\includegraphics*[width=.49\columnwidth]{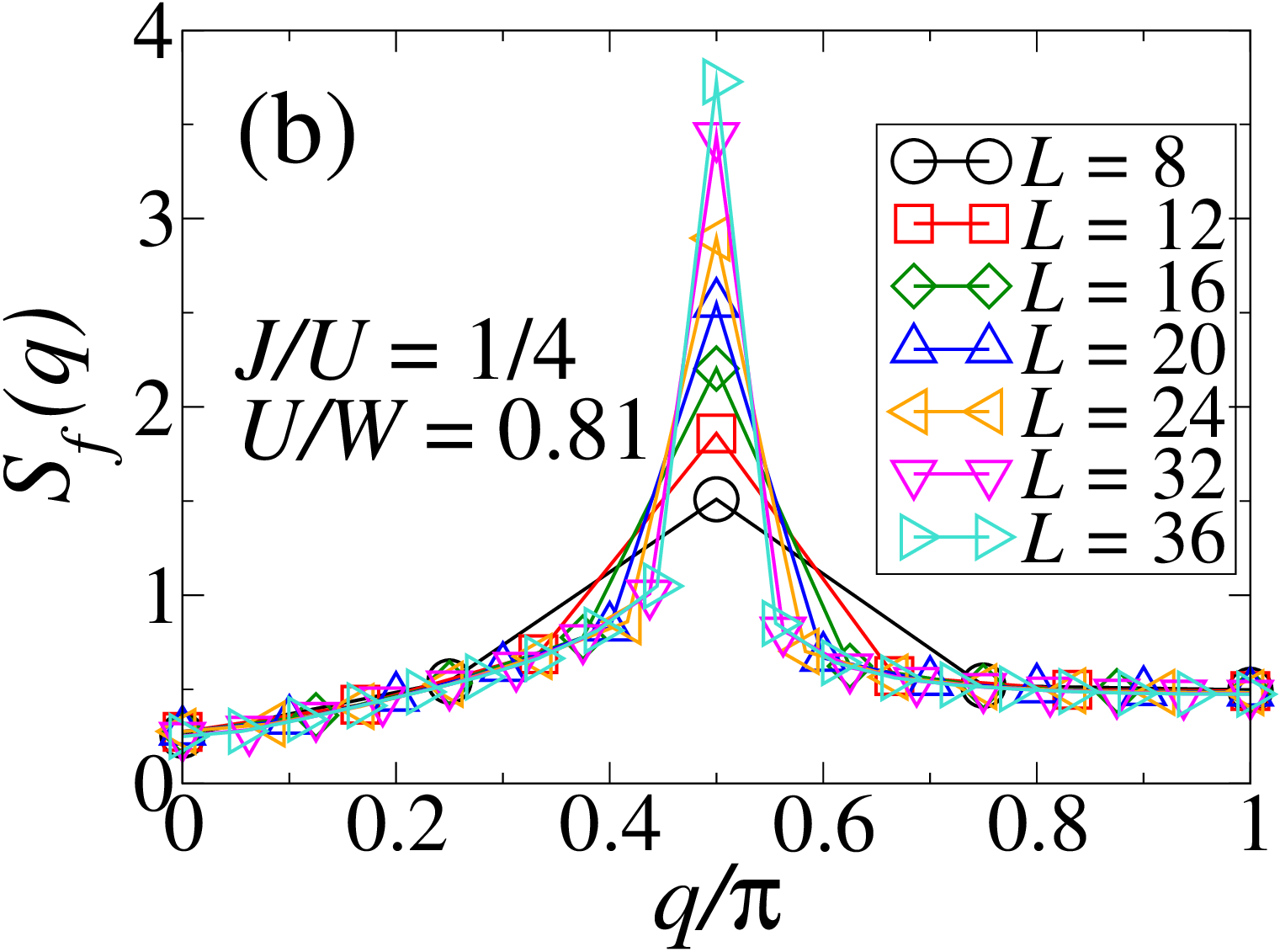}
\includegraphics*[width=.49\columnwidth]{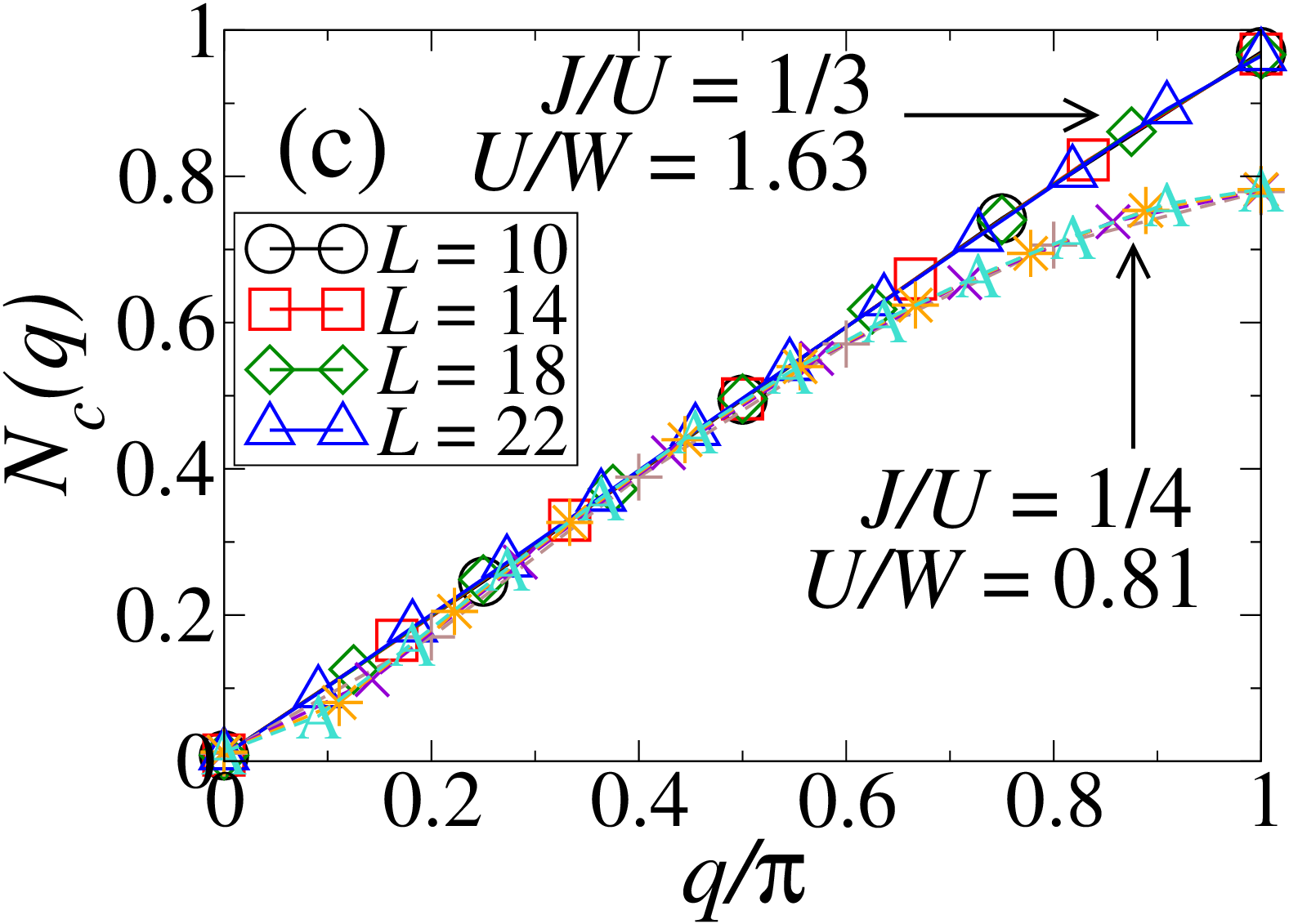}
\caption{(a) Spin correlations in the localized (insulating) band 
vs.\ the normalized distance, at $J/U = 1/4$, $U/W=0.81$, and $L=36$. The formation of ferromagnetic clusters interacting antiferromagnetically is clearly shown. 
(b) Spin structure factor for panel (a) at several $L$'s (symbols). 
A clear peak at $q/\pi=1/2$ is shown. 
(c) Charge structure factor of the itinerant 
bands for the $J/U$ and $U/W$ indicated, and varying $L$ (symbols). 
}
\label{fig:3}
\end{figure}

While previous mean-field studies of the OSMP arrive 
to conclusions qualitatively similar to those
of Fig.~\ref{fig:2}, the DMRG method can reveal 
the fine details of the OSMP spin and charge 
arrangements. For example,
Fig.~\ref{fig:3}(a) contains representative 
OSMP spin-spin correlation functions~\cite{suppl}. Surprisingly, an unexpected pattern 
of spins $\uparrow \uparrow \downarrow \downarrow$ is clearly observed.
The spin structure factor in Fig.~\ref{fig:3}(b) displays 
a sharp peak at $q=\pi/2$, that increases with $L$
(the peak is also robust away from $n = 4/3$). 
Then, at intermediate couplings, our DMRG results 
indicate that exotic Block spin states can be stabilized within the OSMP 
(region ``B'' in Fig.~\ref{fig:1}). 
These Block spin states are qualitatively similar 
to those reported for pnictides and 
selenides~\cite{bao,cao,luo-vacancies,Yu11,huang-mou,dong,BNL,RMP,Dai,caron1,caron2,luo-ladders}.

In addition, a surprising FM OSMP region with a maximal 
value of the spin has been found (see SM~\cite{suppl}).
Its charge structure factor, $N_c(q)$, is shown in Fig.~\ref{fig:3}(c) 
at several $L$'s. Although there are no signs 
of charge order, there is a clear evidence of spinless fermions 
behavior with momentum $q=\pi$. The effective filling of the emergent
spinless fermions can be understood by considering that there are
three electrons per site in the itinerant bands (and one in the localized band).
This is equivalent to one hole per site per two orbitals or, equivalently,
1/2 hole per site on an effective chain of length $2L$. This is a
``half-filled'' spinless fermion system, inducing 
the momentum of $\pi$ in $N_c(q)$. For the B phase, $N_c(q)$ shows a qualitatively 
similar behavior, no charge order, and a spectrum broadening
as shown in Fig.~\ref{fig:3}~(c)~\cite{suppl}.

\begin{figure}
\includegraphics*[width=.75\columnwidth]{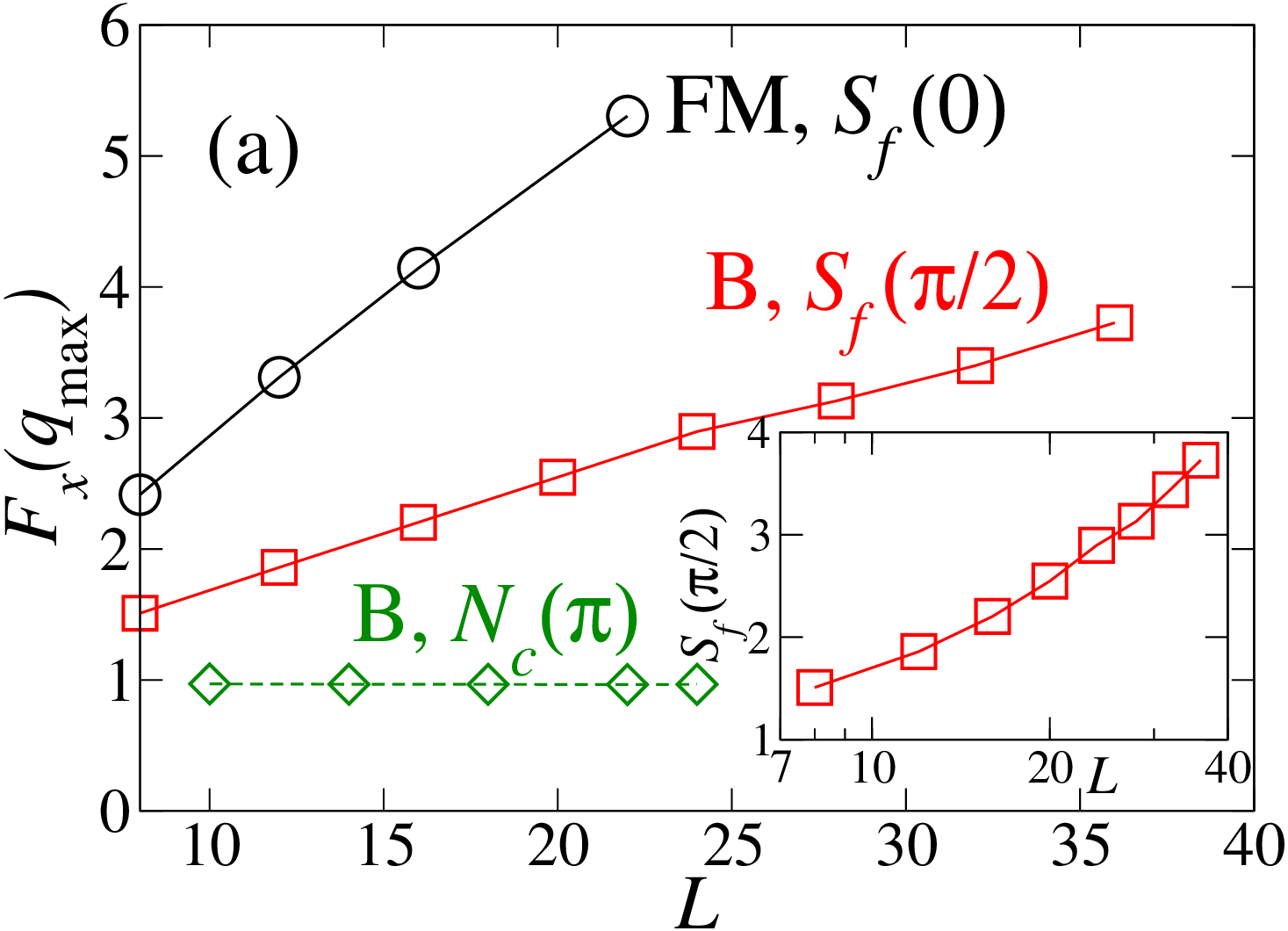}
\includegraphics*[width=.80\columnwidth]{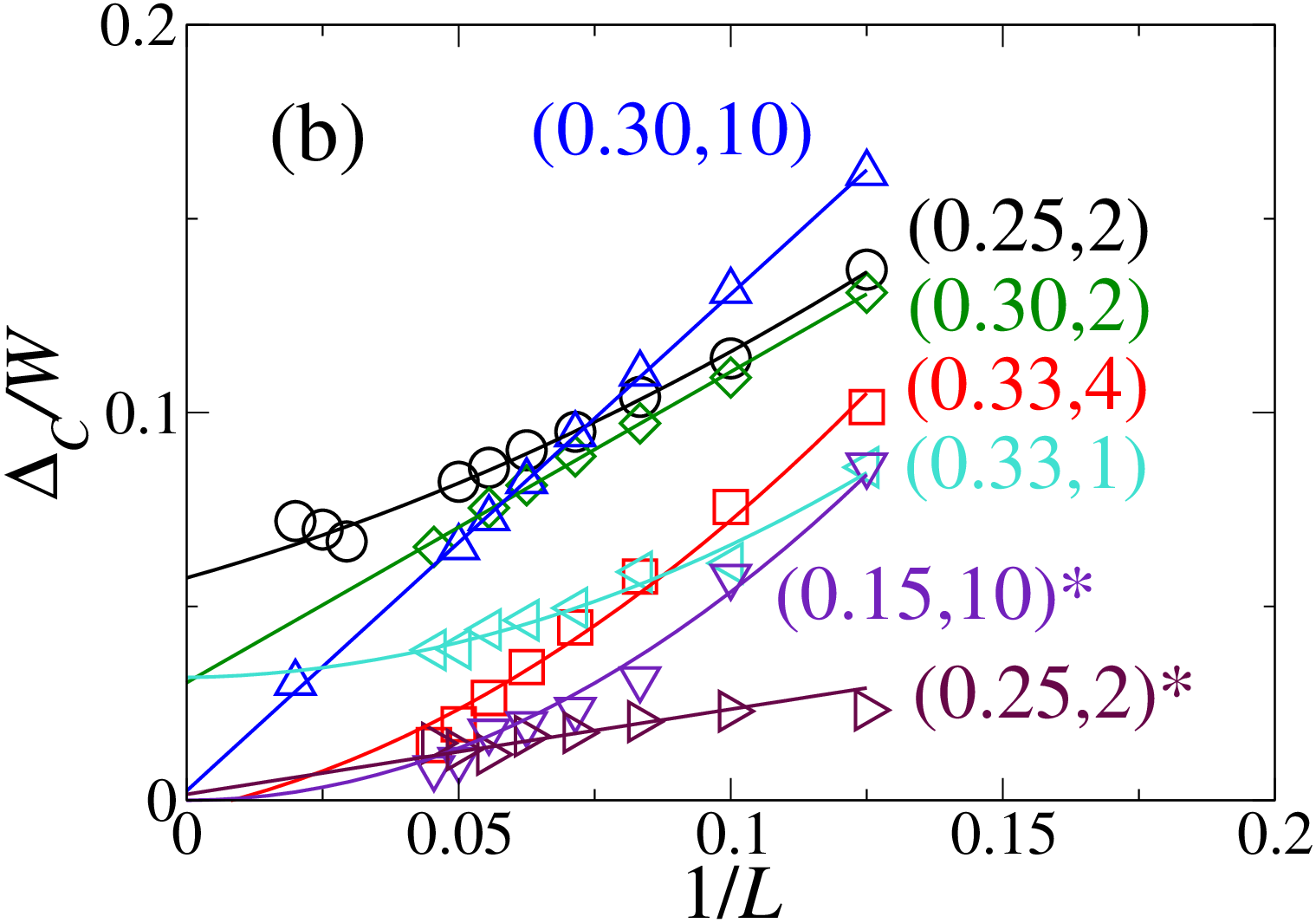}
\caption{(a) Size dependence of the peaks of the charge (dashed line) and spin (solid line) structure factors for the FM and B phases with couplings $J/U=1/4$, $U/W=0.81$ and $J/U=1/3$, $U/W=1.63$, respectively. \textit{Inset}: Semi-logarithmic plot of $S_f(\pi/2)$. 
(b) Charge gap vs.\ $1/L$ for the values of the pair $(J/U,U)$ shown. Results with the ``*'' symbol stand for the 2-hole doped case.
}
\label{fig:4}
\end{figure}

Figure~\ref{fig:4}(a) shows the finite-size dependence of the peak of 
the charge and spin structure factors, generically referred to as $F_{x}(q_{\rm max})$, for the 
B and FM regions. $F_x$ becomes $S_f$ 
and $N_c$ for spin and charge, respectively, at the maximal 
momentum $q_{\rm max}$. The FM state displays a quasi-linear 
increase of the structure factor peak vs.~$L$ signaling a quasi-long 
range order in the spin correlations. The B phase presents 
a logarithmic increase with $L$ indicating, again, (quasi-long-ranged) 
power-law behavior of the spin correlations with non-trivial exponents 
[see inset Fig.~\ref{fig:4}(a)]. On the other hand, the charge peak 
shows no order tendencies in any of the phases evidencing only short-ranged 
correlations; however, the charge correlations show a non-trivial 
trend associated with spinless behavior, as discussed in Fig.~\ref{fig:3}(c).

The metallic vs.~insulating nature of the OSMP is difficult to address, but
guidance can be obtained by studying the charge gap $\Delta_C$~\cite{suppl} 
[see Fig.~\ref{fig:4}(b)]. 
While the extrapolation to the bulk is difficult, ``by eye'' it is clear 
that the extrapolations appear to converge 
to small gaps in the $0.05\times W$ to $0.00$ range. Moreover, 
upon light 2-hole doping the gap vanishes. Thus, a likely scenario is that the Block states 
at exactly $n=4/3$ have a small gap, that rapidly closes with doping. Since in the realistic five-orbital
Hubbard model for pnictides the combined population of the $xz$, $yz$, and $xy$ orbitals is not exactly 4,
the relevant OSMP regime would be metallic.

{\it Relation to Kondo lattice models.} The results reported here unveil 
interesting analogies with previous studies of the Double Exchange (DE) models 
for Mn-oxides~\cite{dagotto01}. This is natural, since DE physics relies on the interplay of
localized and itinerant degrees of freedom, as it occurs in the OSMP regime.  
This DE-OSMP relation is rigorous:
for the case of a two-band model, it has been shown that the low-energy effective theory of the OSMP
is related to a FM Kondo lattice model (FMKLM) with Hubbard repulsion in the 
conduction band~\cite{Biermann05}. For a generic multiorbital case, one can show 
that the effective model turns into a ``correlated'' FMKLM. If the system has $n_{\rm orb}$ orbitals, 
this model describes the FM exchange between the localized band with magnetic moments 
and a many-body bath of itinerant electrons. The bath is 
built out of the remaining $n_{\rm orb}-1$ 
orbitals and the only interaction between localized and itinerant bands 
is the FM exchange.

This DE-OSMP relation provides further support to the main conclusions of our effort.
In fact, previous DMRG and Monte Carlo investigations of the DE model 
already unveiled a variety of so-called ``island phases'' 
containing small clusters of various shapes made of ferromagnetically
aligned spins and with AFM couplings among 
them, both in one and two dimensions~\cite{island1,island12,island13,island2}. 
These phases are qualitatively the analog of the Block phases, 
although for quite different couplings, electronic densities, and band structures.
The previous DE results~\cite{island1,island12,island13,island2}
reinforces our conjecture that the Block states observed here 
via the detailed analysis of examples are likely stable 
for a wider variety of models, both 
in one and two dimensions. Since in manganites it was shown 
that exotic non-FM phases need the crucial addition of a superexchange 
coupling $J_{\rm AFM}$ between the localized 
spins~\cite{dagotto01,yunoki1,yunoki2},
it is natural to conjecture that a similar coupling must be incorporated into
the effective model in the OSMP regime of iron superconductors and related
compounds.

{\it Conclusions.}
In this publication, multiorbital Hubbard models have been investigated using
DMRG techniques in one dimension, and the phase diagrams have been constructed. 
Our main result is that the OSMP
regime contains unexpected internal structure. In particular, phases with
the characteristics of Block states (FM clusters, AFM coupled) have been 
identified. Full FM has also been found within the OSMP. The Block states likely
arise from the frustration generated 
by competing ferromagnetic/antiferromagnetic 
tendencies in the model, as it happens with the island phases in manganites.
Block phases were discussed before in models of Fe-based superconductors, 
but were not associated with the OSMP regime. Moreover, since 
a formal mapping between the Hubbard
and Kondo lattice models exists in the OSMP regime, 
our results establish an unexpected 
analogy between Double-Exchange physics, where similar
Block phases were reported before, and that of multiorbital
models for iron superconductors. 
Our conclusions are of direct relevance to
real quasi one-dimensional materials but they may apply
to two-dimensional systems as well, considering that ``island
phases'' in the two dimensional Kondo models have been 
reported before~\cite{island2}.

Support by the Early Career Research Program, 
Scientific User Facilities
Division, Basic Energy Sciences, US Department of Energy,
under contract with UT-Battelle (J.R. and G.A.) and by the 
U.S. Department of Energy, Office of Basic Energy Sciences, 
Materials Sciences and Engineering Division (J.R.) is acknowledged.
For this project A.M. and E.D. were supported by the 
National Science Foundation
under Grant No. DMR-1104386.


\begin{thebibliography}{10}

\bibitem{Anisimov} V. Anisimov, I. Nekrasov, D. Kondakov, T. M. Rice, and M. Sigrist, Eur. Phys. J. B {\bf 25}, 191 (2002).

\bibitem{Vojta} M. Vojta, J. Low. Temp. Phys. \textbf{161}, 203 (2010).

\bibitem{Georges13} A. Georges, L. de' Medici, and J. Mravlje, Annu. Rev. Cond. Mat. Phys. \textbf{4}, 137 (2013).

\bibitem{Liebsch04} A. Liebsch, Phys. Rev. B \textbf{70}, 165103 (2004).

\bibitem{Biermann05} S. Biermann, L. de' Medici, and A. Georges, Phys. Rev. Lett. \textbf{95}, 206401 (2005).

\bibitem{Medici09} L. de' Medici, S. R. Hassan, M. Capone, and X. Dai, Phys. Rev. Lett. \textbf{102}, 126401 (2009).


\bibitem{Ishida10} H. Ishida and A. Liebsch, Phys. Rev. B \textbf{81}, 054513 (2010).

\bibitem{Medici11b} L. de' Medici, Phys. Rev. B \textbf{83}, 205112 (2011).

\bibitem{Yu12} R. Yu and Q. Si, Phys. Rev. B {\bf 86}, 085104 (2012).

\bibitem{Yu13} R. Yu and Q. Si, Phys. Rev. Lett. \textbf{110}, 146402 (2013).

\bibitem{Medici-new} L. de' Medici, G. Giovannetti, and M. Capone, arXiv:1212.3966.

\bibitem{Bascones12} E. Bascones, B. Valenzuela, and M. J. Calder\'on, 
Phys. Rev. B \textbf{86}, 174508 (2012).

\bibitem{Valenzuela13} B. Valenzuela, M. J. Calder\'on, 
G. Le\'on, E. Bascones, Phys. Rev. B \textbf{87}, 075136 (2013).

\bibitem{Liebsch11} A. Liebsch, Phys. Rev. B \textbf{84}, 180505(R) (2011).

\bibitem{Yi-2012} M. Yi, D. Lu, R. Yu, S. Riggs, J.-H. Chu, B. Lv, Z. Liu, M. Lu, Y. Cui, M. Hashimoto, S.-K. Mo, Z. Hussain, C.-W. Chu, 
I. Fisher, Q. Si, and Z.-X. Shen, Phys. Rev. Lett. {\bf 110}, 067003 (2013).

\bibitem{lanata} N. Lanat\'a, H. U. R. Strand, G. Giovannetti, B. Hellsing, 
L. de' Medici, and M. Capone, Phys. Rev. B {\bf 87}, 045122 (2013).

\bibitem{Greger13} M. Greger, M. Kollar, and D. Vollhardt, Phys. Rev. Lett. \textbf{110}, 046403 (2013).

\bibitem{bao} W. Bao, Q. Huang, G. F. Chen, M. A. Green, D. M. Wang, J. B. He,
X. Q. Wang, and Y. Qiu, Chin. Phys. Lett. {\bf 28}, 086104 (2011).

\bibitem{cao} C. Cao, and J. Dai, Phys. Rev. Lett. {\bf 107}, 056401 (2011).

\bibitem{luo-vacancies} Q. Luo, A. Nicholson, J. Riera, D.-X. Yao, 
A. Moreo, and E. Dagotto, Phys. Rev. B {\bf 84}, 140506(R) (2011).

\bibitem{Yu11} R. Yu, J.-X. Zhu, and Q. Si, Phys. Rev. Lett. {\bf 106},
186401 (2011).

\bibitem{huang-mou} S.-M., Huang and C.-Yu Mou, Phys. Rev. B {\bf 84}, 184521 (2011).

\bibitem{dong} W. Li, S. Dong, C. Fang, and J.-P. Hu, Phys. Rev. B {\bf 85}, 100407(R) (2012).

\bibitem{BNL} W.-G. Yin, C.-C. Lee, and W. Ku, Phys. Rev. Lett. {\bf 105}, 107004 (2010).

\bibitem{RMP} E. Dagotto, Rev. Mod. Phys. {\bf 85} 849 (2013), and references therein.

\bibitem{Dai} P. Dai, J.P. Hu , and E. Dagotto, Nat. Physics {\bf 8}, 709 (2012).

\bibitem{caron1} J. M. Caron, J. R. Neilson, D. C. Miller, A. Llobet, and T. M.
McQueen, Phys. Rev. B {\bf 84}, 180409(R) (2011).

\bibitem{caron2} J. M. Caron, J. R. Neilson, D. C. Miller, K. Arpino, 
A. Llobet, and T. M. McQueen, Phys. Rev. B {\bf 85}, 180405(R) (2012).

\bibitem{luo-ladders} Q. L. Luo, A. Nicholson, J. Rinc\'on, S. Liang, J. Riera, G. Alvarez,
L. Wang, W. Ku, G. Samolyuk, A. Moreo, and E. Dagotto, 
Phys. Rev. B {\bf 87}, 024404 (2013).

\bibitem{dmrg1} S. R. White, Phys. Rev. Lett. {\bf 69}, 2863 (1992); \emph{ibid.} {\bf 48}, 10345 (1993).

\bibitem{dmrg2} U. Schollw\"ock, Rev. Mod. Phys. {\bf 77}, 259 (2005). 

\bibitem{dmrg3} K. Hallberg, Adv. Phys. {\bf 55}, 477 (2006).

\bibitem{daghofer} M. Daghofer, A. Nicholson, A. Moreo, and E. Dagotto, Phys. Rev. B {\bf 81}, 014511 (2010).




\bibitem{suppl} The supplementary material 
at [URL will be inserted by publisher] contains details of the 
definition of physical quantities, and additional numerical results.

\bibitem{rong-comment} This BI+M phase (not to be confused with a phase separated state
involving the BI and M states) has characteristics similar to those reported at small $J$
in R. Yu and Q. Si, Phys. Rev. B {\bf 84}, 235115 (2011). Further studies 
beyond the scope of
the present publication can clarify its properties, as well as the true
existence of a sharp transition between the M and BI+M states.






\bibitem{Haule} K. Haule and G. Kotliar, New J. of Phys. {\bf 11}, 025021 (2009).

\bibitem{luo-neutrons} Q. Luo, G. Martins, D.-X. Yao, M. Daghofer, R. Yu, 
A. Moreo, and E. Dagotto, Phys. Rev. B {\bf 82}, 104508 (2010).

\bibitem{Kotliar} Z. P. Yin, K. Haule, and G. Kotliar, Nat. Mater. {\bf 10}, 932 (2011).

\bibitem{valenti} J. Ferber, K. Foyevtsova, R. Valent\'i, 
and H.O. Jeschke, Phys. Rev. B {\bf 85}, 094505 (2012). 
See also Y.-Z. Zhang, Hunpyo Lee, H.-Q. Lin, C.-Q. Wu, H. O. Jeschke, and R. Valent\'i,  
Phys. Rev. B {\bf 85}, 035123 (2012), and references therein.


\bibitem{footnotedelta} The inset of Fig.~S6 in the SM
shows that $n_0$ and $n_1$ are very
close in the OSMP regime, but not identical. The same occurs in Fig.~\ref{fig:2} (a). The
reason is that the initial small 
energy splitting between levels $0$ and $1$ is washed out in the
OSMP where $U/W$ is of order 1 or larger.




\bibitem{dagotto01} E. Dagotto, T. Hotta, and A. Moreo, Physics Reports {\bf 344}, 1 (2001).

\bibitem{island1} D. J. Garc\'ia, K. Hallberg, 
C. D. Batista, M. Avignon, and B. Alascio, Phys. Rev. Lett. {\bf 85}, 3720 (2000).

\bibitem{island12} D. J. Garcia, K. Hallberg, C. D. Batista, S. Capponi, 
D. Poilblanc, M. Avignon, and B. Alascio, Phys. Rev. B {\bf 65}, 134444 (2002).

\bibitem{island13} D. J. Garcia, K. Hallberg, B. Alascio, and M. Avignon,
Phys. Rev. Lett. {\bf 93}, 177204 (2004).

\bibitem{island2} H. Aliaga, B. Normand, K. Hallberg, M. Avignon, and B. Alascio,
Phys. Rev. B {\bf 64}, 024422 (2001).  

\bibitem{yunoki1} S. Yunoki and A. Moreo, Phys. Rev. B {\bf 58}, 6403 (1998).

\bibitem{yunoki2} A. Moreo, M. Mayr, A. Feiguin, S. Yunoki, and E. Dagotto,
Phys. Rev. Lett. {\bf 84}, 5568 (2000).


\end{thebibliography}
\end{document}


\title{Supplementary Material for\\ Exotic Magnetic Order in the Orbital-Selective Mott Regime of Multiorbital Systems}

\author{Juli\'an Rinc\'on}
\affiliation{Center for Nanophase Materials Sciences, Oak Ridge National Laboratory, Oak Ridge, Tennessee 37831, USA}
\affiliation{Materials Science and Technology Division, Oak Ridge National Laboratory, Oak Ridge, Tennessee 37831, USA}

\author{Adriana Moreo}
\affiliation{Materials Science and Technology Division, Oak Ridge National Laboratory, Oak Ridge, Tennessee 37831, USA}
\affiliation{Department of Physics and Astronomy, The University of Tennessee, Knoxville, Tennessee 37996, USA}

\author{Gonzalo Alvarez}
\affiliation{Center for Nanophase Materials Sciences, Oak Ridge National Laboratory, Oak Ridge, Tennessee 37831, USA}
\affiliation{Computer Science \& Mathematics Division, Oak Ridge National Laboratory, Oak Ridge, Tennessee 37831, USA}

\author{Elbio Dagotto}
\affiliation{Materials Science and Technology Division, Oak Ridge National Laboratory, Oak Ridge, Tennessee 37831, USA}
\affiliation{Department of Physics and Astronomy, The University of Tennessee, Knoxville, Tennessee 37996, USA}

\date{\today}

\maketitle


In this section, 
technical details concerning the numerical calculation and lattice models are provided.
Also supplementary results regarding other states in the phase diagrams of the two 
Hamiltonians analyzed here are presented.

\section{Implementation of the DMRG technique}

The Density Matrix Renormalization Group (DMRG) technique~\cite{dmrg1s,dmrg2s,dmrg3s}
was used in this publication, with $L$ up to 50 sites.
The number of states per block kept during the DMRG iterations were 900
and up to 19 sweeps were performed during the finite-size algorithm evolution,
giving rise to a discarded weight of the order $10^{-6}-10^{-4}$.
For small system sizes, $L=4$ and 6, DMRG results
were confirmed by contrast against an exact diagonalization.
Finite-size effects are surprisingly small in the phase diagrams reported
here~\cite{Sakamoto02}.
The full open source code, sample input decks, and corresponding 
computational details have been made available 
at \url{https://web.ornl.gov/~gz1/papers/49/}.

\begin{figure}
\includegraphics*[width=0.85\columnwidth]{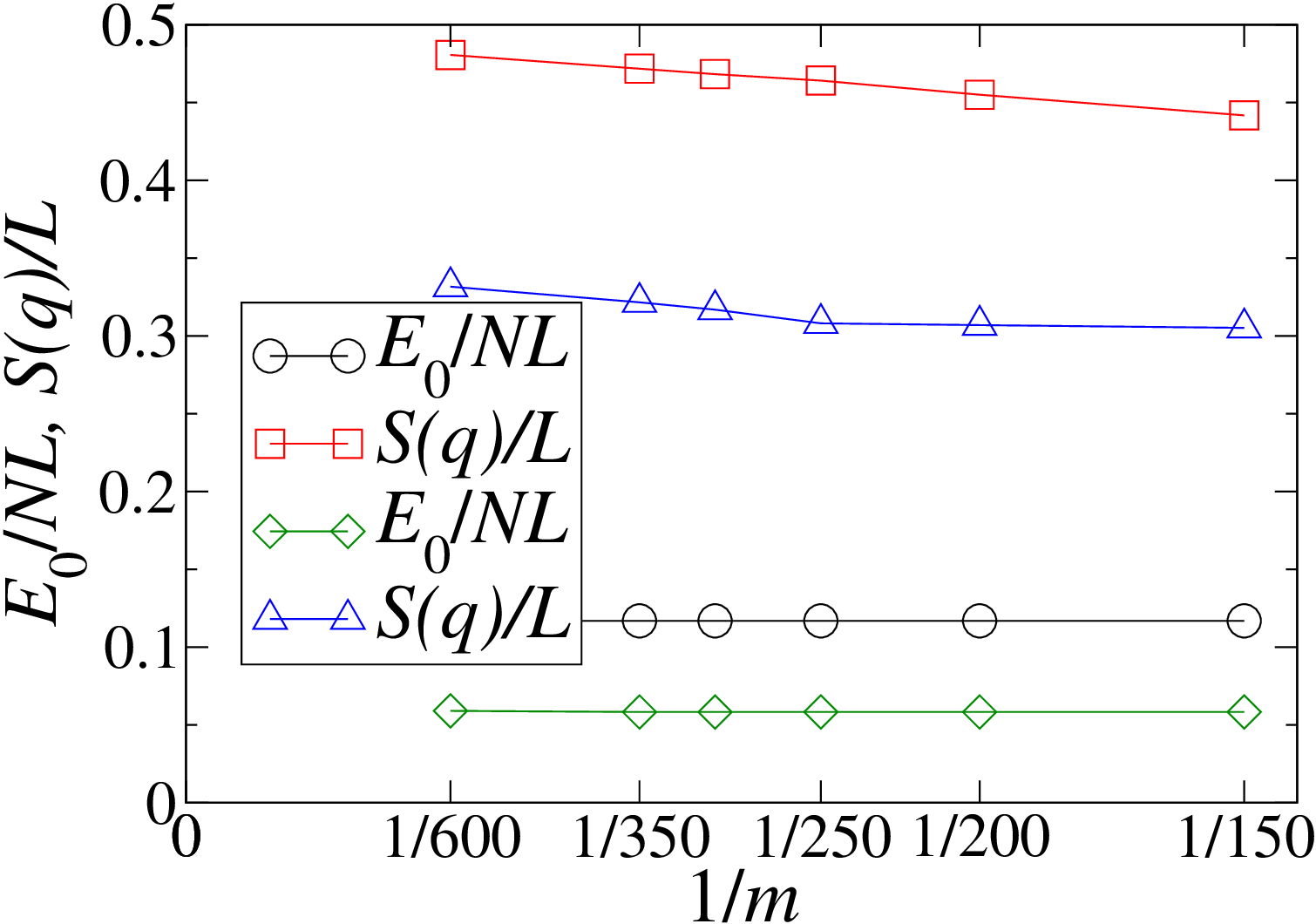}
\caption{Dependence of some observables with the number of states kept in the
DMRG calculations, $m$. Shown are the ground state energy (per site and per number of
electrons) and the peak height (per site) of the spin structure factor in
the Block phase (evaluated at $q = \pi/2$). Circles and squares are results 
for $L=12$, while diamonds and triangles are for $L=24$. All these results are
for $U/W=0.81$, $J/U=0.25$, and $n=4/3$.
}
\label{fig:error}
\end{figure}

 To address the accuracy of the DMRG calculations performed in this work,
Fig.~\ref{fig:error} displays the ground state energies and the height of the peaks
of the spin structure factor in the Block phase as a function of the number
of states kept, $m$, for $L=12$ (symbols: circles and squares) and 24
(symbols: diamonds and triangles). Mean values such
as the ground state energy show a smooth, almost flat, dependence
with $m$, as expected~\cite{dmrg2s,dmrg3s}. This quantity has clearly reached 
convergence for the number of states used for both lengths, $L$, considered. 
However, correlation functions typically have a more complex evolution 
with the number of states kept.
By performing quadratic fits to the data, it was observed that the error in the
heights of the peaks for 12 and 24 sites correspond to $4\%$ and $15\%$,
respectively. This level of error is acceptable 
given the computational complexity of the three-orbital model worked out here.
It should be remarked that within DRMG the study of a three-orbital Hubbard model on
a chain of length, say, $L=36$ and with $m=800$ states is approximately equivalent to the
study of a one-orbital Hubbard model using $L=108$ and $m=12800$. This illustrates that
the numerical effort presented here is substantial.

\section{Definition of Model~2 and results}

\begin{figure}
\includegraphics*[width=.761\columnwidth]{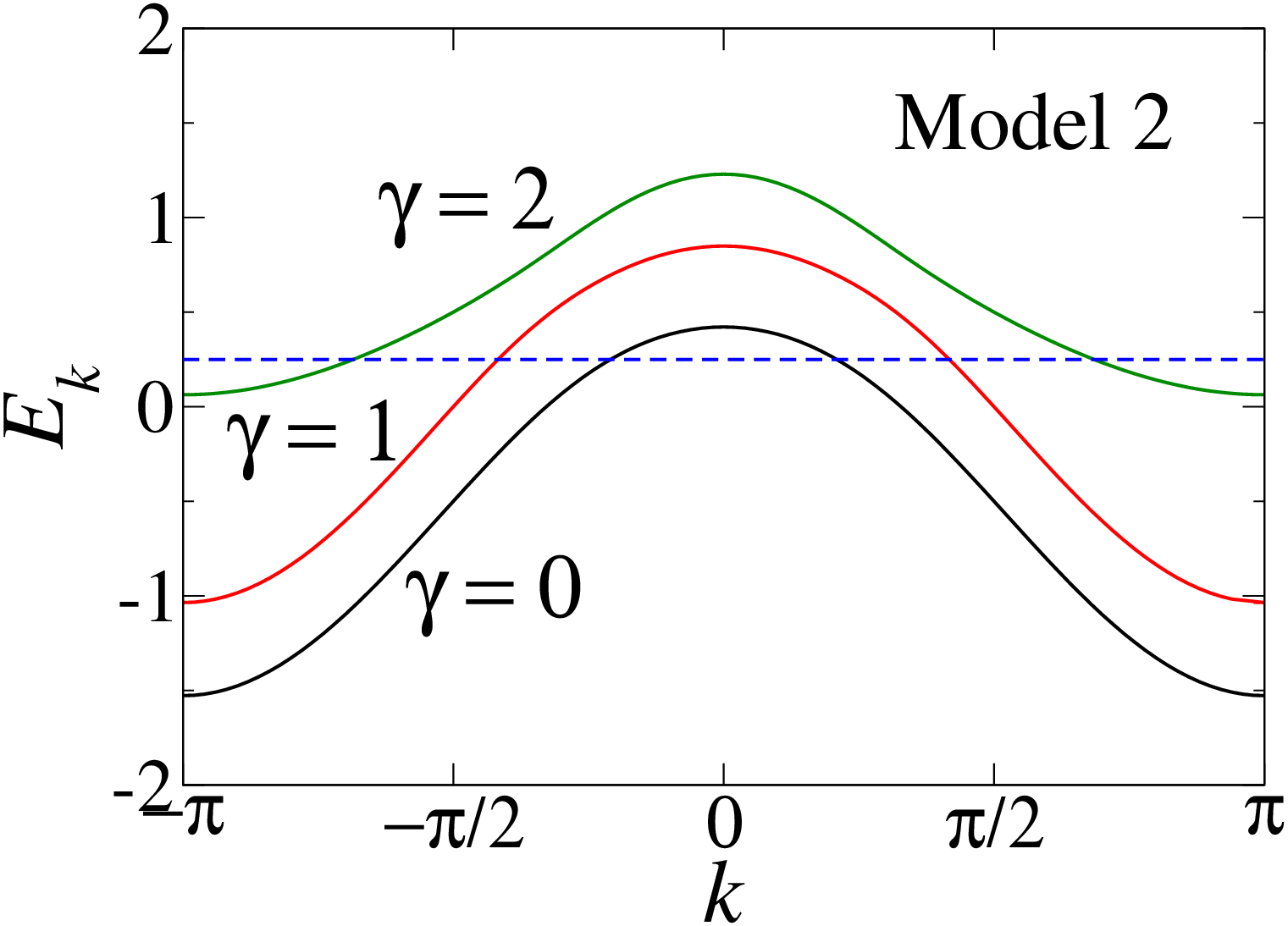}
\caption{The band structure of Model~2.  The blue (dashed) line is the Fermi energy at $n=4/3$.
The bandwidth is $W=5.5|t_{00}|$. The difference with Model~1 is mainly in the size of the hole pockets.} 
\label{fig:S1}
\end{figure}

To help determine how general the conclusions reported in the main text are, in the present
effort two model Hamiltonians, that slightly differ in their band structures as shown in 
Fig.~\ref{fig:S1} and in the hoppings of Table~S1, have been studied.
These models were named Model~1 and Model~2. 
Although both of these
models still contain a combination of hole and electron pockets,
their main differences are in the size of the hole pockets. 
In both cases, the 
filling factor will be fixed to $n=4/3$ (i.e., four particles 
every three orbitals), in analogy with the $n$ used 
for three-orbital models for iron superconductors, as discussed in the main text.
Results for Model~1 were presented in the main text, while results for Model~2 
are presented here. 

\begin{table}[b]
\caption{\label{tab:t1} Set of parameters that 
define the models studied in this work. Hoppings and
$\Delta_\gamma$ were defined in the main text.}
\begin{ruledtabular}
\begin{tabular}{ccccccc}
& $t_{00}=t_{11}$ & $t_{02}$ & $t_{12}$ & $\Delta_0$ & $\Delta_1$ & $\Delta_2$ \\ \hline
Model~1 & $-0.5$ & $-0.15$ & 0.1 & $-0.1$ & 0.0 & 0.8\\
Model~2 & $-0.5$ & $-0.25$ & 0.1 & $-0.5$ & 0.0 & 0.5\\
\end{tabular}
\end{ruledtabular}
\end{table}

\begin{figure}
\includegraphics*[width=.49\columnwidth]{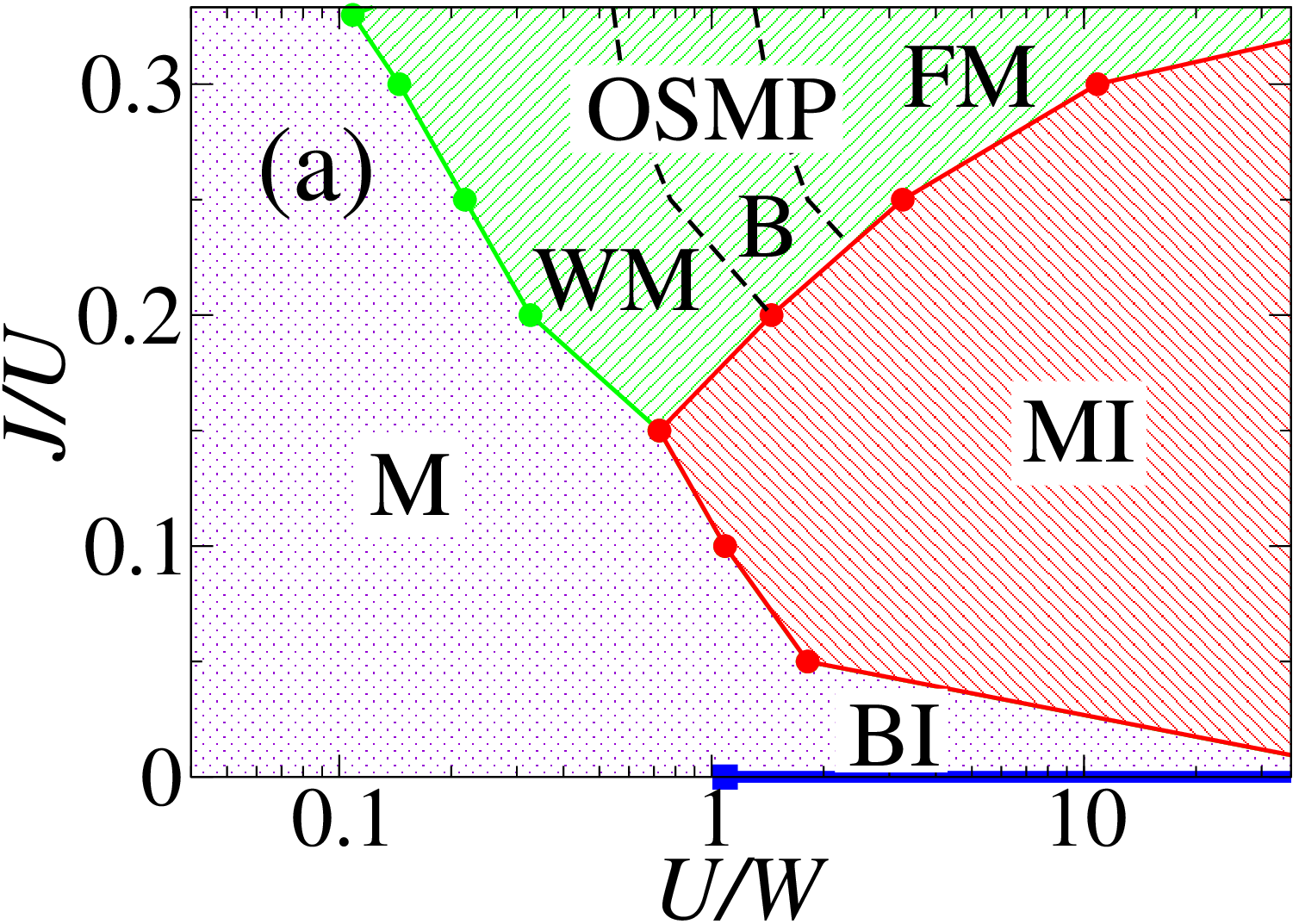}                                                                       
\includegraphics*[width=.49\columnwidth]{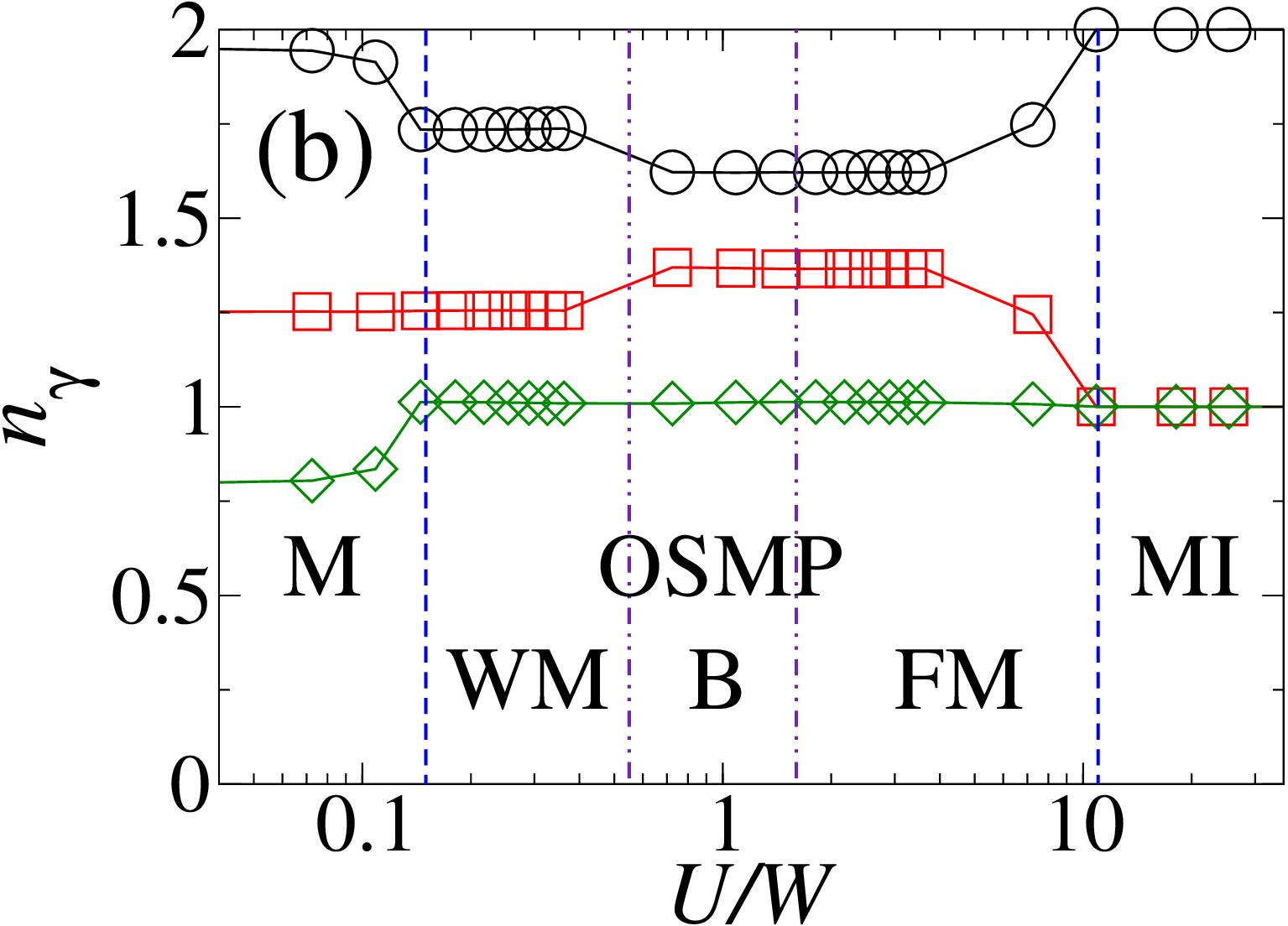}                                                                       
\includegraphics*[width=\columnwidth]{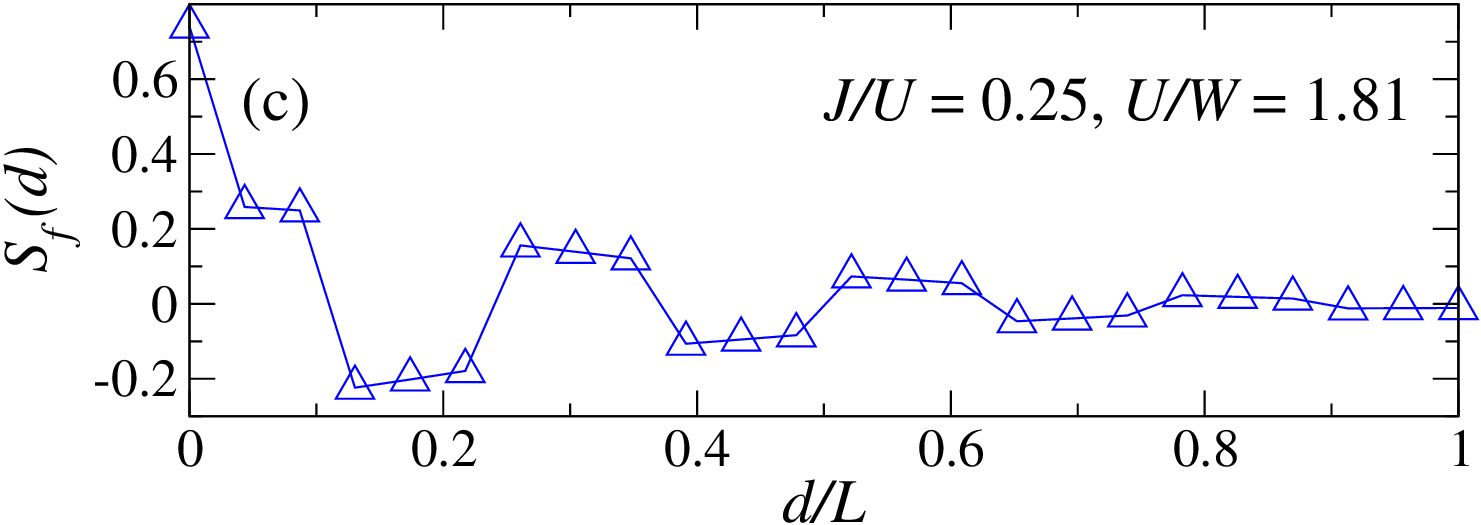}                                                                          
\caption{Results for Model~2: (a) Phase diagram at $n = 4/3$. Within the OSMP there is an additional phase characterized by the presence of weak magnetism (WM).
Other labels are as already defined in the main text for Model~1.      
(b) Orbital occupancy, $n_\gamma$, vs.\ $U/W$, for $J/U=0.30$.                                                           
The OSMP is found within the range $0.15 < U/W < 10$. (c) Spin correlations in the localized (insulating) band           
vs.\ distance for $J/U=0.25$, $U/W=1.81$, and $L=24$ sites, illustrating the formation                                   
of the Block state $\uparrow \uparrow \uparrow \downarrow \downarrow \downarrow$.   
}
\label{fig:S2}
\end{figure}

Figure~\ref{fig:S2} contains the DMRG results for Model~2. Panel (a) has the phase diagram while
Panel (b) displays the orbital occupations, to illustrate how the phase diagram was constructed.                         
There are clear similarities with the results for Model~1, since the M, MI, and OSMP regimes are                         
present. Within OSMP, there is a Block (B) region as well,                                                               
although in this case the spin pattern [that can be deduced from Panel (c)]                                              
is $\uparrow \uparrow \uparrow \downarrow \downarrow \downarrow$. The Fermi surface differences                          
presumably lead to the B-region different spin patterns between Models 1 and 2.
The FM regime within the OSMP is also present in Model~2, as it is in Model~1.
However, the BI+M regime is absent, and in the OSMP there is a new regime 
with weak magnetization in Model~2.                      
In spite of these small differences, the general issue addressed in this                                              
publication, namely the existence of the OSMP and the presence of Block phases inside, is common                         
to both Models 1 and 2, suggesting that these results could be generic                                                   
beyond the models used here~\cite{warning}.

\section{Definition of Observables}

The several phases appearing in our models have been identified by calculating expectation values and two-point 
correlation functions. For instance, the occupation number of each orbital, defined as
\begin{equation}
n_\gamma = \frac{1}{L}\sum_{i,\sigma} \langle n_{i \sigma \gamma}\rangle,
\end{equation}
has been calculated, as well as their respective magnetizations,
\begin{equation}
S^z_\gamma = \frac{1}{2L}\sum_{i} \langle n_{i \uparrow \gamma}-n_{i \downarrow \gamma}\rangle.
\end{equation}
With regards to correlation functions, the focus has been on the full spin and charge correlations defined as follows:
\begin{equation}
n_{i,j} = \langle n_{i} n_{j} \rangle,
\end{equation}
and
\begin{equation}
S_{i,j} = \langle \mathbf{S}_{i } \cdot \mathbf{S}_{j} \rangle,
\end{equation}
where $n_i = \sum_{\gamma} n_{i \gamma}$ and $\mathbf{S}_{i}= \sum_{\gamma} \mathbf{S}_{i \gamma}$.
Since there is a close relationship between the OSMP state and the ground state of a Kondo lattice model~\cite{Tsun97}, 
as discussed in the main text, the charge correlations of the itinerant bands have also been evaluated,
\begin{equation}
N_{i,j}^c = \langle n_{i }^c n_{j }^c \rangle,
\end{equation}
as well as the spin correlations of the Mott-insulating orbital, 
\begin{equation}
S_{i,j}^f = \langle \mathbf{S}_{i}^f \cdot \mathbf{S}_{j}^f \rangle.
\end{equation}
The ``itinerant'' charge was defined as 
\begin{equation}
n_{i}^c = \sum_{\gamma=0,\,1} n_{i \gamma},
\end{equation}
and the ``localized'' spin as 
\begin{equation}
\mathbf{S}_{i}^f = \sum_{\alpha,\beta} c_{i \beta 2}^+ \bm{\tau}_{\alpha \beta} c_{i \beta 2},
\end{equation}
with $\bm{\tau}$ being the Pauli matrices. These quantities are defined \emph{in the spirit} of a Kondo lattice model calculation~\cite{Tsun97}.

With $n$ being the mean value of a generic charge operator $\mathcal{N}_{j}$, 
the corresponding structure factors for charge and spin are defined as
\begin{equation}
\mathfrak{N}(q) = \frac{1}{L}\sum_{k,j} e^{-i q (j-k)} \langle (\mathcal{N}_{k} - n)( \mathcal{N}_{j} - n) \rangle,
\label{Nq}
\end{equation}
and
\begin{equation}
\mathfrak{S}(q) = \frac{1}{L}\sum_{k,j} e^{-i q (j-k)} \langle \mathcal{S}_{k} \cdot \mathcal{S}_{j} \rangle,
\label{Sq}
\end{equation}
respectively. If in Eq.~(\ref{Nq}) the generic charge 
operator is $\mathcal{N}_{i} = n_i$,
then the structure factor $N(q)$ is obtained. If instead the definition 
$\mathcal{N}_{i} = n_i^c$ is used, then $N_c(q)$ is obtained. 
If in Eq.~(\ref{Sq}), the total spin operator 
$\mathcal{S}_{i} = \mathbf{S}_i$
and the ``localized'' spin operator $\mathcal{S}_{i} = \mathbf{S}^f_i$ are inserted, then the 
corresponding Fourier transforms $S(q)$ and $S_f(q)$, respectively, are obtained. For the case of spin operators, 
the spatial spin correlations shown in the main text are calculated as the average over their respective 
correlations at a fixed distance $d$~\cite{Xavier10}, namely,
\begin{equation}
S(d) = \frac{1}{L-d}\sum_{j=1}^{L-d} \langle \mathbf{S}_j \cdot \mathbf{S}_{j+d} \rangle.
\end{equation}

The charge gap is given as the difference of energies as
\begin{equation}
\Delta_C = E_0(N+1) + E_0(N-1) - 2E_0(N),
\end{equation}
where $E_0(N)$ is identified as the ground state energy of the system with $N$ particles.

\section{Additional Results for Model~1}

\begin{figure}
\includegraphics*[width=.85\columnwidth]{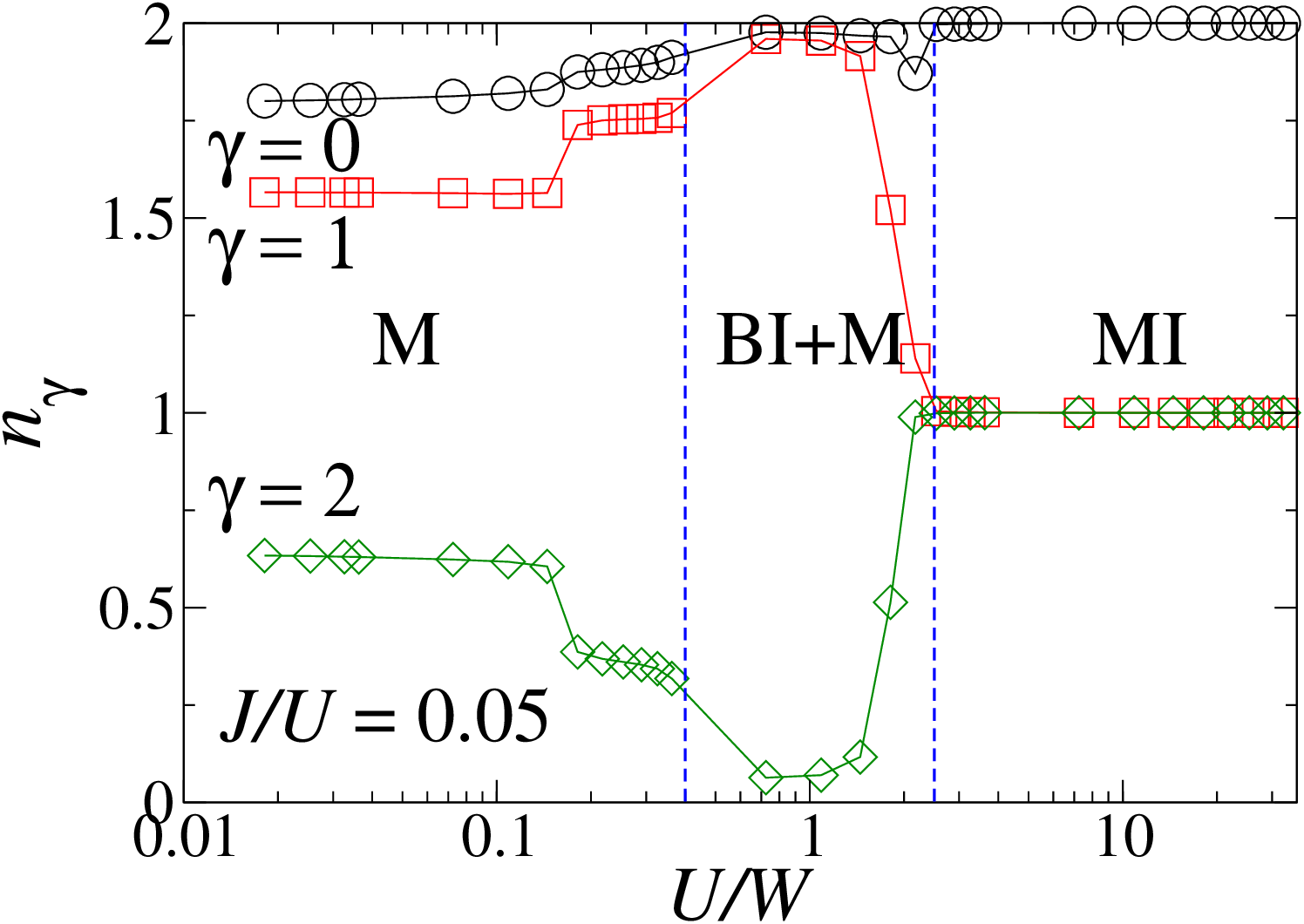}
\caption{Mean value of the occupancy corresponding to orbital $\gamma$, $n_\gamma$, 
as a function of the local repulsion $U$ in units of 
the bandwidth $W$, at a fixed $J/U = 0.05$. 
The phase BI+M is present in the range $0.4 < U/W < 2$. The model used is Model~1}
\vspace{-0.2cm}
\label{fig:S3}
\end{figure}

In this section, additional results are presented corresponding 
to the FM phase of the OSMP regime and also for the BI+M phase observed 
at lower values of $J/U$ and in the intermediate range $U/W\sim 1$, 
since both states were not analyzed in detail in the main text. 
Only additional results for Model~1 are presented in this supplementary material,
but similar results were found for Model~2.

Figure~\ref{fig:S3} shows the particle occupation number 
at each orbital vs.\ $U/W$, at a fixed (relatively small) $J/U = 0.05$ and $L=24$ sites. 
At small $U/W<0.4$, a metallic region is expected with 
each orbital being partially occupied. In the other extreme, i.e. 
in the strong coupling regime ($U/W>4$), clear signatures of 
a full Mott insulating phase are found with all the occupations being integers. 
In the intermediate regime of couplings, 
a phase that can be considered as a slightly doped band insulator was observed.
This phase was called BI+M since there is a clear dominance of a ``correlated'' band insulator state
(finite $U/W$) together with metallic fluctuations. Here, the orbital occupations are not integers but
close to integers.

\begin{figure}
\includegraphics*[width=.85\columnwidth]{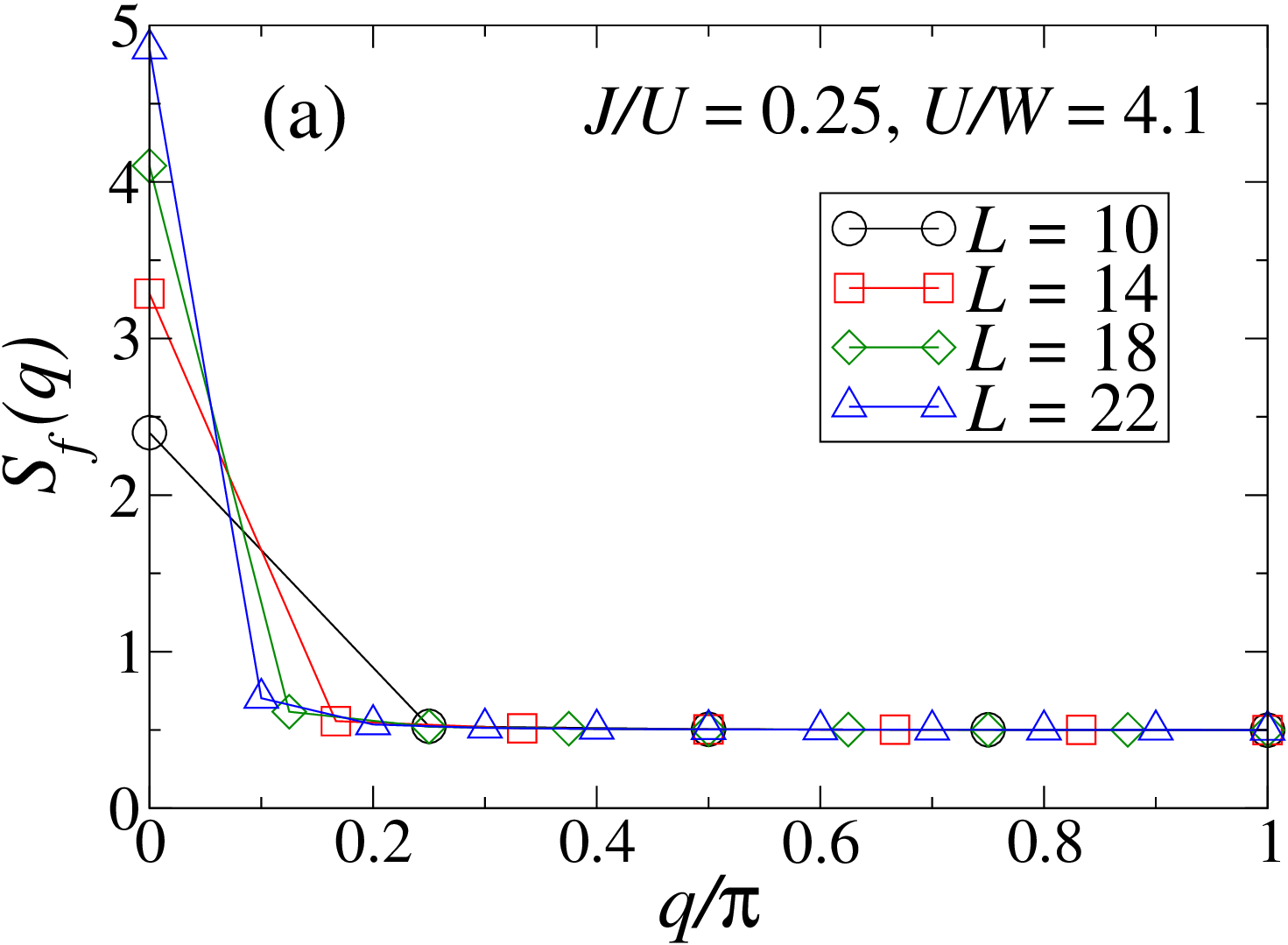}
\includegraphics*[width=.85\columnwidth]{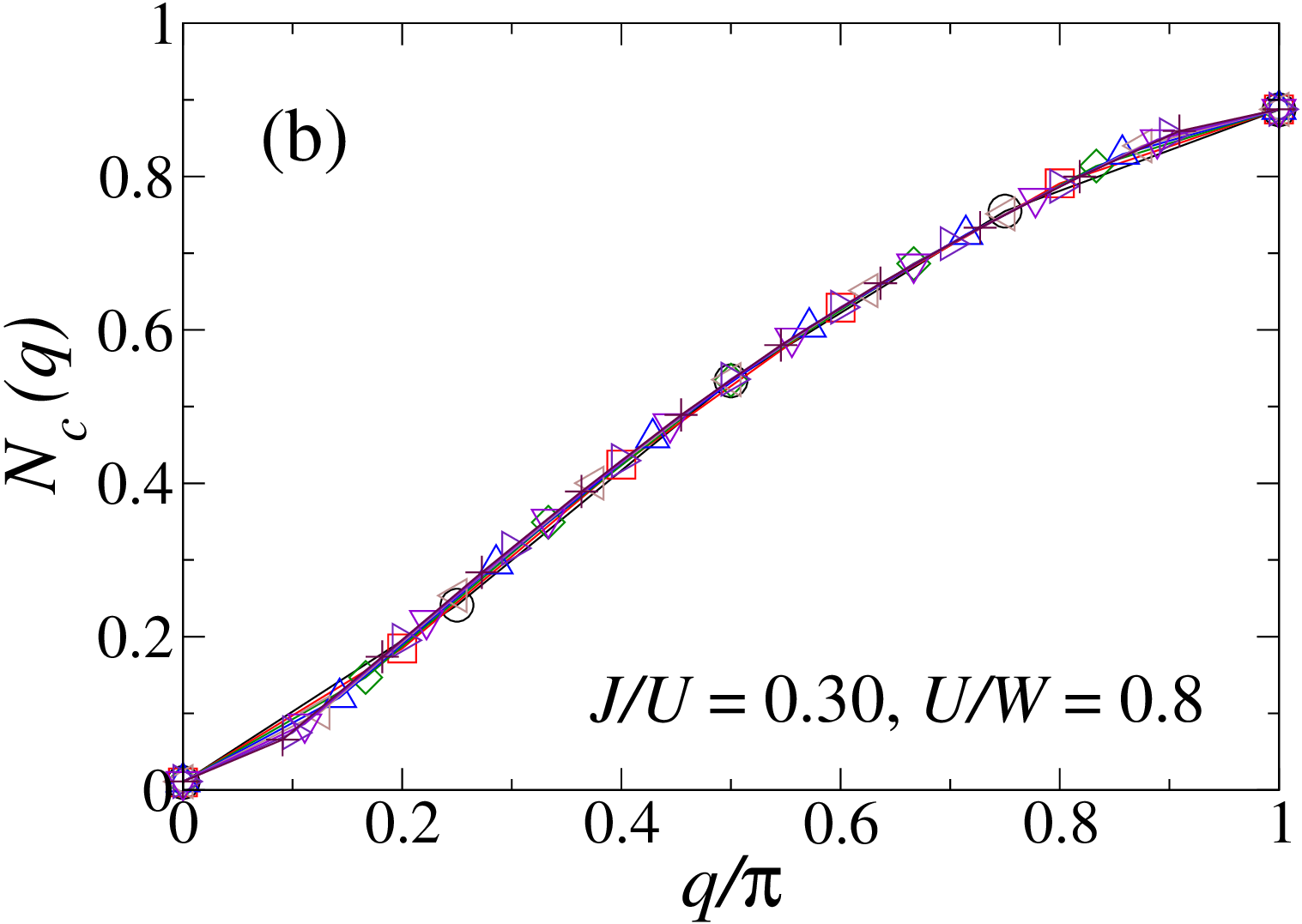}
\caption{Localized-band spin and itinerant charge structure factors in the (a) FM and (b) B phases, 
respectively, for different values of the couplings $(J/U,U/W)$ and system sizes. 
For $S_f(q)$, clear signatures of robust FM are found. For $N_c(q)$, a broad momentum 
distribution is observed. The model used is Model~1.}
\vspace{-0.2cm}
\label{fig:S4}
\end{figure}

In Fig.~\ref{fig:S4}~(a), the ``localized'' spin structure factor, $S_f(q)$, for the FM phase is shown 
for different lattice sizes at fixed coupling constants, $J/U = 1/4$ and $U/W = 4.1$. As can be 
observed from the figure, a robust peak at $q=0$ develops indicating the presence of FM order within the OSMP state.
Note that the height of this peak increases as $L$ increases, suggesting a dominant 
power-law decaying long-range magnetic order
in this one dimensional system, as discussed in the main text.
The spatial correlation function (not shown) also displays clear FM features, as expected.

In Fig.~\ref{fig:S4}~(b), the ``itinerant'' charge structure factor, $N_c(q)$, is shown for 
several system lengths, at $J/U = 0.3$ and $U/W = 0.8$. As can be observed 
in the phase diagram of Model~1, this choice of parameters corresponds to the Block (B) magnetic phase 
in the OSMP regime. The magnetic structure factor of this 
phase has already been discussed in the main text. As shown in Fig.~\ref{fig:S4} (b), 
the charge in the itinerant bands within the OSMP regime
corresponds to that of spinless fermions with a broad momentum distribution, as also
discussed in the main text. 
This behavior is in agreement with previous results reported in 
numerical studies of the Kondo lattice model~\cite{island1s,island12s,island13s}.

\begin{figure}
\includegraphics*[width=.85\columnwidth]{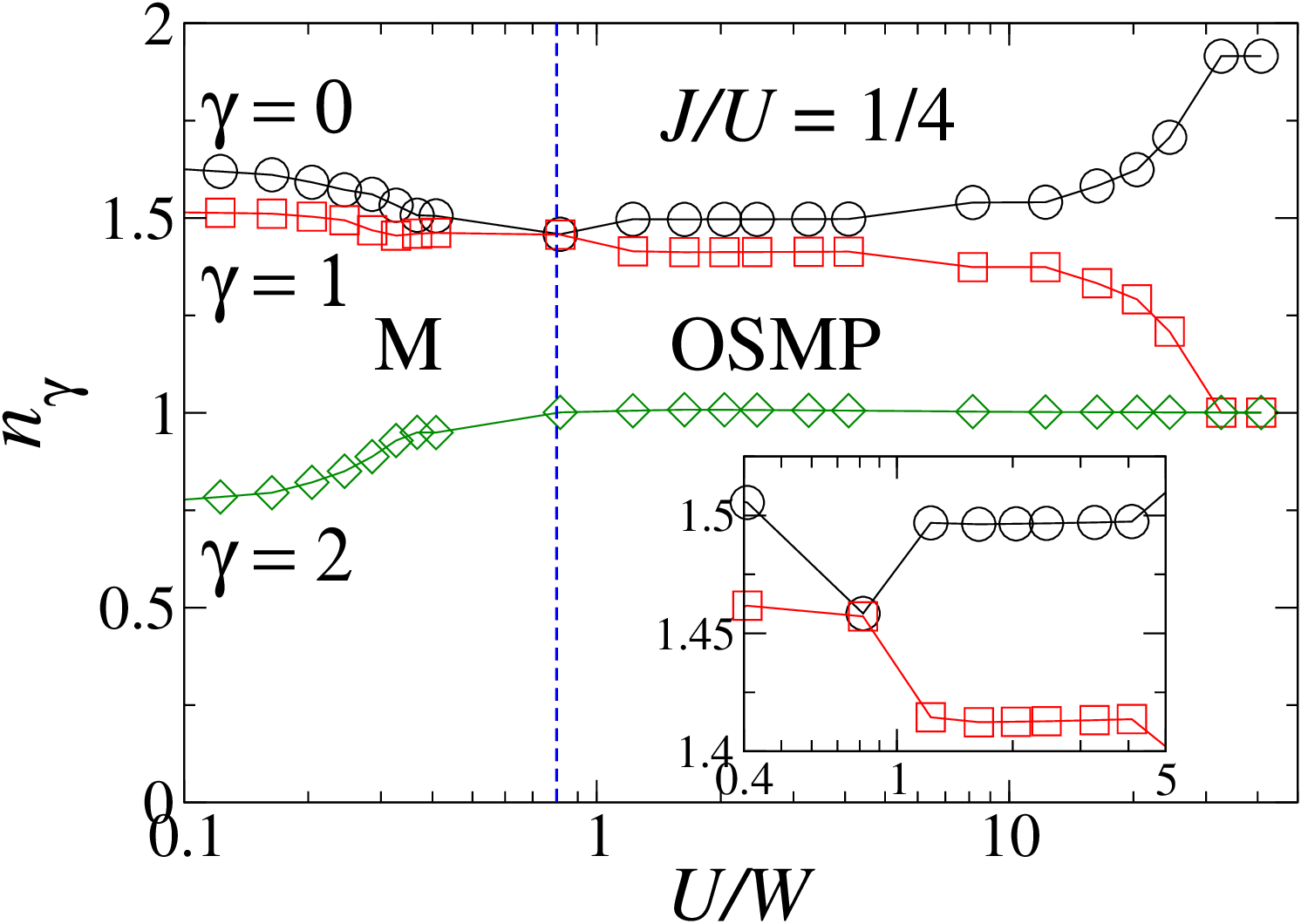}
\caption{Results for Model~1: Mean value of the electronic occupancy, 
$n_\gamma$, corresponding to orbital $\gamma$
vs.~the local repulsion $U/W$ in bandwidth $W$ units at $J/U = 1/4$. The total electronic density corresponds to 2 holes added to
the system, as compared with the results shown in Fig.3 (a) for
the undoped $n = 4/3$ case.
Inset: detail of $n_\gamma$ vs.\ $U/W$ in the OSMP regime. }
\label{fig:S5}
\end{figure}

Finally, in Fig.~\ref{fig:S5} the Model~1 orbital occupation is shown for a case identical
to that presented in the main text for $n=4/3$, but with two holes added. The purpose
of this study is to confirm that the electronic densities $n_1 \sim n_2 \sim 1.5$ observed
in Fig.~3 (main text) change with doping while $n_2$ remains equal to 1.0 in the OSMP
regime. As shown in Fig.~\ref{fig:S5}, indeed this occurs suggesting that the
regime observed in our study is compatible with the properties of the OSMP. Moreover,
the inset also shows that the electronic densities $n_0$ and $n_1$ while very close
in value are not identical. In the OSMP regime that starts at $U/W \sim 1$, 
the effect of the original energy splitting
between orbitals 0 and 1 diminishes as $U/W$ further grows 
and these orbitals become quasi degenerate in the OSMP.